\documentclass[acmsmall,screen]{acmart}

\usepackage{listings}
\usepackage{xcolor}
\usepackage{tcolorbox}
\lstdefinelanguage{diff}{
  morecomment=[f][\color{gray}][0]{@@},
  morecomment=[f][\color{red}]{-},
  morecomment=[f][\color{green!50!black}]{+},
  morecomment=[f][\color{blue}]{---}, 
  morecomment=[f][\color{blue}]{+++},
}
\usepackage{etoolbox}
\AtBeginEnvironment{tcolorbox}{\footnotesize}

\usepackage{subcaption}
\usepackage{pifont}
\usepackage{enumitem}
\usepackage{graphicx}
\usepackage{wrapfig}
\usepackage[framemethod=tikz]{mdframed}
\mdfdefinestyle{customstyle}{
  linecolor=gray!100,
  linewidth=3pt,
  innerleftmargin=3pt,
  topline=false,
  rightline=false,
  bottomline=false,
  leftline=true,
  innerrightmargin=3pt,
  innertopmargin=3pt,
  innerbottommargin=3pt,
  backgroundcolor=gray!15
}

\newenvironment{custommdframed}
  {\begin{mdframed}[style=customstyle]}
  {\end{mdframed}}

\begin{document}

\title{An Empirical Study on Failures in Automated Issue Solving}

\author{Simiao Liu}
\affiliation{%
  \institution{Beihang University}
  \city{Beijing}
  \country{China}
}
\email{buaalsm@buaa.edu.cn}

\author{Fang Liu}\authornote{Corresponding Author}
\email{fangliu@buaa.edu.cn}
\affiliation{%
  \institution{Beihang University}
  \city{Beijing}
  \country{China}
}

\author{Liehao Li}
\affiliation{%
  \institution{Beihang University}
  \city{Beijing}
  \country{China}
}

\author{Xin Tan}
\affiliation{%
  \institution{Beihang University}
  \city{Beijing}
  \country{China}
}

\author{Yinghao Zhu}
\affiliation{%
  \institution{The University of Hong Kong}
  \city{Hong Kong}
  \country{China}
}

\author{Xiaoli Lian}
\affiliation{%
  \institution{Beihang University}
  \city{Beijing}
  \country{China}
}

\author{Li Zhang}
\affiliation{%
  \institution{Beihang University}
  \city{Beijing}
  \country{China}
}

\begin{abstract}
Automated issue solving seeks to autonomously identify and repair defective code snippets across an entire codebase. SWE-Bench has emerged as the most widely adopted benchmark for evaluating progress in this area. While Large Language Model (LLM)-based agentic tools show great promise, they still fail on a substantial portion of tasks. Moreover, current evaluations primarily report aggregate issue-solving rates, which obscure the underlying causes of success and failure, making it challenging to diagnose model weaknesses or guide targeted improvements. 
To bridge this gap, we first analyze the performance and efficiency of three state-of-the-art tools, spanning both pipeline-based and agentic architectures, in automated issue solving tasks of SWE-Bench-Verified (a validated subset of SWE-Bench) under varying task characteristics. Furthermore, to move from high-level performance metrics to underlying cause analysis, we conducted a systematic manual analysis of 150 failed instances. From this analysis, we developed a comprehensive taxonomy of failure modes comprising 3 primary phases, 9 main categories, and 25 fine-grained subcategories. Then we systematically analyze the distribution of the identified failure modes, the results reveal distinct failure fingerprints between the two architectural paradigms, with the majority of agentic failures stemming from flawed reasoning and cognitive deadlocks. Motivated by these insights, we propose a collaborative Expert-Executor framework. It introduces a supervisory \textit{Expert} agent tasked with providing strategic oversight and course-correction for a primary \textit{Executor} agent. This architecture is designed to correct flawed reasoning and break the cognitive deadlocks that frequently lead to failure. Experimental results show that our framework successfully resolves 22.2\% of previously intractable issues for a state-of-the-art single agent. These findings pave the way for building more robust agents through diagnostic evaluation and collaborative design. We provide the replication package at \url{https://anonymous.4open.science/r/IssueSolvingEmpirical}.
\end{abstract}





\maketitle

% FSE: At the time of submission, each paper should have no more than 18 pages for all text and figures, plus 4 pages for reference

\section{Introduction}
%需要先介绍issue solving的backgroud
Automated issue resolution has become a critical challenge in both software engineering and artificial intelligence fields. Its objective is to autonomously identify and repair defective code snippets across entire codebases, guided by reported issues. Developers typically devote the majority of their debugging effort to understanding code and implementing the necessary modifications ~\cite{alaboudi2023constitutes}. 
Recent advances in Large Language Models (LLMs) \cite{wang2025reuse,zheng2025towards,wang2025agents} and LLM-based agentic tools have significantly advanced the automation of software development tasks. By leveraging their strong capabilities in code understanding and generation, LLMs demonstrate promising performance in tasks such as code generation \cite{guo2024deepseekcoder,hui2024qwencoder}, fault localization \cite{qin2024agentfl,widyasari2024FuseFL}, program repair \cite{xia2024chatrepair,zhang2025patch}, and automated issue solving \cite{zhang2024autocoderover,xia2024agentless}.

%However, despite these advances, the field faces a critical bottleneck. Current evaluation practices are a primary source of this problem. The community has largely focused on aggregate pass/fail rates on benchmarks like SWE-Bench \cite{jimenez2024swe-bench} and SWE-Bench-Verified \cite{SWE-Bench-Verified} which treat agents as "black boxes". 

SWE-Bench \cite{jimenez2024swe-bench} stands out as the most popular benchmark for automated issue solving, comprising 2,294 tasks from 12 well-maintained Python repositories. Each task is paired with an issue description and the corresponding repository snapshot where the issue is to be resolved. 
The tool to be evaluated is required to generate a patch to address the given issue, and the generated patch is evaluated by running tests associated with the issue.
OpenAI further creates SWE-Bench-Verified \cite{SWE-Bench-Verified}, which contains 500 verified issues from SWE-bench ensuring clear problem specifications and consistent evaluation environments.
This benchmark provides a standardized and realistic setting for assessing the strengths and limitations of LLMs in repository-level issue resolution. 
Its public leaderboard\footnote{\url{https://www.swebench.com/index.html}} has become a focal point for competition, attracting top solutions from leading AI companies like Anthropic\footnote{\url{https://www.anthropic.com/}} and Cognition AI\footnote{\url{https://cognition.ai/}}. This race to the top has steadily driven up reported issue-solving success rates, marking significant progress in automated issue solving.

Despite these advances, LLM agents still fail on a substantial portion of SWE-Bench tasks. For example, OpenHands + CodeAct v2.1 (Claude-3.5 Sonnet) \cite{neubig2024-openhands-codeact-2.1:} achieved a 53.00\% issue-solving rate, which represented state-of-the-art open-source solution as of February 2, 2025. The more critical issue, however, is that current evaluation practices, which focus primarily on aggregate solving rates, provide little insight into the nature and underlying causes of these failures. Such aggregate issue-solving rates treat agents as black boxes, telling us they fail, but not \textit{how} or \textit{when} of the problem-solving process. This lack of diagnostic information makes it difficult for researchers to pinpoint specific weaknesses in agent reasoning or tool interaction, and thus hinders targeted improvements. Without a deeper understanding of the failure modes \cite{liu2024exploring}, further progress risks becoming inefficient.
Besides, prior work \cite{yao2023reactsynergizingreasoningacting, gao2024agentscope} has primarily focused on boosting performance by optimizing prompts, tool integrations, or scaffolding strategies. In contrast, much less attention has been paid to understanding concrete failures or examining how these failures vary across different architectural paradigms.

To bridge this critical gap, we conduct the first in-depth empirical study of failure modes in state-of-the-art automated issue solving tools. Our study, grounded in the OpenAI-vetted SWE-Bench-Verified benchmark, is framed by four research questions:

\textbf{RQ1 (Performance and Efficiency Analysis)} examines the performance and efficiency of three representative tools (OpenHands \cite{wang2024openhands}, Tools Calude \cite{Tools-Claude}, and Agentless \cite{xia2024agentless}) in automated issue solving task on SWE-Bench-Verified, where all these tools embody the dominant architectural paradigms and were among the top performers on the SWE-Bench leaderboard at the time of our study. This RQ aims to uncover not only whether these tools are successful, but also \emph{how} their performance and efficiency diverge under varying issue characteristics, setting the stage for a deeper diagnosis. Our analysis reveals that, while the tools achieve comparable overall issue-solving rates, they exhibit complementary strengths. Performance consistently declines as task complexity increases, especially for multi-file modification tasks. Moreover, agentic tools frequently become trapped in long, unproductive loops.

To move from high-level performance metrics to underlying cause analysis, \textbf{RQ2 (A Taxonomy of Failure Modes)} develops a structured taxonomy of failure modes of the studied tools in RQ1, encompassing 9 primary failure categories and 25 fine-grained subcategories, organized across three issue-solving phases: Location, Repair, and Iteration \& Validation. This taxonomy is derived from a detailed manual analysis of 150 issues where at least one studied tool failed.

\textbf{RQ3 (Characterization of Failure Modes)} further examines how these failure modes are distributed across tools and task difficulty levels, uncovers their underlying root causes, and assesses their downstream impacts on the issue-solving process. Our analysis reveals that different tools exhibit distinct failure patterns: pipeline-based tools tend to fail early, often due to incorrect problem localization, whereas agent-based tools more often get stuck in repetitive loops during the later repair stage. The most common cause of failures is fundamental flaws in the agent's reasoning, which lead it to persist with a flawed plan.

The findings from RQ3 motivates \textbf{RQ4 (Mitigation Exploration with a Collaborative Architecture)}, where we propose \textbf{Expert--Executor} model, a collaborative architecture to address the observed failure patterns. It emulates the review processes of human development teams. The evaluation results on 108 issues previously failed by the baseline \texttt{OpenHands} agent show that our collaborative architecture successfully resolves 24 of these challenging cases, achieving performance comparable to top-tier single-agent systems that leverage more powerful proprietary models, such as Claude 4 Sonnet. 
The improvements are most pronounced in mitigating failures during the \textbf{Iterative Verification} stage, such as \textit{bug reproduction failures} \cite{mukherjee2021fixing} and \textit{iteration anomaly failures}. Furthermore, it can prevent strategy defects in the \textbf{Repair} phase, such as \textit{evasive repair}.
These results provide strong evidence that incorporating an explicit review mechanism substantially improves agents' strategic reasoning and self-correction throughout the automated issue-solving lifecycle.

In summary, this paper makes the following contributions:

\begin{itemize}[leftmargin=*]
    \item \textbf{Systematic Empirical Study.} We conduct the first in-depth empirical study contrasting pipeline-based and agent-based tools on SWE-Bench-Verified, revealing their distinct performance characteristics and architecture-specific vulnerabilities.

    \item \textbf{In-depth Failure Analysis and Insights.} We introduce a comprehensive taxonomy for LLM agents' failures in automated issue solving. Leveraging this taxonomy, our fine-grained analysis of 150 failure cases provides the first qualitative characterization of why and how these tools fail, pinpointing root causes like early-stage diagnostic errors in pipeline-based tools and unproductive iterative loops in agent-based tools.

    \item \textbf{Failure Mitigation Strategy.} Inspired by our empirical findings, we propose a collaborative Expert-Executor framework to address the observed failures, paving the way for more robust and cooperative AI-driven software development.

    \item \textbf{A Publicly Available Dataset of Annotated Failures.} We release our manually annotated dataset of 342 failure instances. To our knowledge, this is the first public resource of its kind, providing a valuable foundation for future research in diagnosing, understanding, and mitigating LLM failures in issue solving tasks.
\end{itemize}

\section{Background \& Related Work: Automated Issue Solving}

\subsection{Benchmark}
A critical prerequisite for advancing automated issue resolution is the availability of realistic and rigorous benchmarks that capture the full complexity of real-world software maintenance \cite{renzullo2025automated}. The SWE-Bench \cite{jimenez2024swe-bench} family of benchmarks has emerged as the de facto standard for evaluating repository-level issue resolution, providing an environment that closely mirrors software engineering workflows.

\textbf{SWE-Bench} \cite{jimenez2024swe-bench} comprises 2,294 tasks collected from 12 well-maintained open-source Python repositories, including diverse domains such as scientific computing, web frameworks, and data processing libraries. Each task is defined by three key components: (i) a snapshot of the repository at the time the issue was reported, (ii) the natural language issue description, and (iii) the corresponding ground-truth fix submitted by developers. The goal for an automated framework is to generate a patch that resolves the issue. 
Evaluation is conducted automatically by running the repository's existing test suite. To be accepted, the model must generate a valid patch in the form of a \texttt{git diff}, which can be applied directly to the provided repository snapshot. A solution is considered correct only if the patch applies successfully and all relevant tests pass \cite{Ye_2021}.

Although SWE-Bench is realistic, it contains noise and unsolvable tasks due to underspecified issues, misleading tests, and fragile environments.
To address this, OpenAI released \textbf{SWE-Bench-Verified}~\cite{SWE-Bench-Verified}, a curated set of 500 tasks vetted by professional developers. Each task is triple-annotated to ensure clear specifications, valid tests, and consistent execution environments, with problematic samples removed. The benchmark also provides a Docker-based evaluation harness and per-issue difficulty labels that reflect the estimated human time-to-fix, which we classify as \emph{Easy} (\textless{}15 min), \emph{Medium} (15–60 min), and \emph{Hard} (\(\ge\)1 h). For the \emph{Hard} category, we merge the original 1–4 h and \textgreater{}4 h buckets because the latter (\textgreater{}4 h) is rare in our sample (3 cases). Owing to its reliability and realism, we adopt SWE-Bench-Verified as the basis for our stratified and reproducible analysis.

\subsection{LLM-based Issue Solving}

\begin{figure*}[t]
    \setlength{\abovecaptionskip}{0.1cm}
    \centering
    \includegraphics[width=0.9\linewidth]{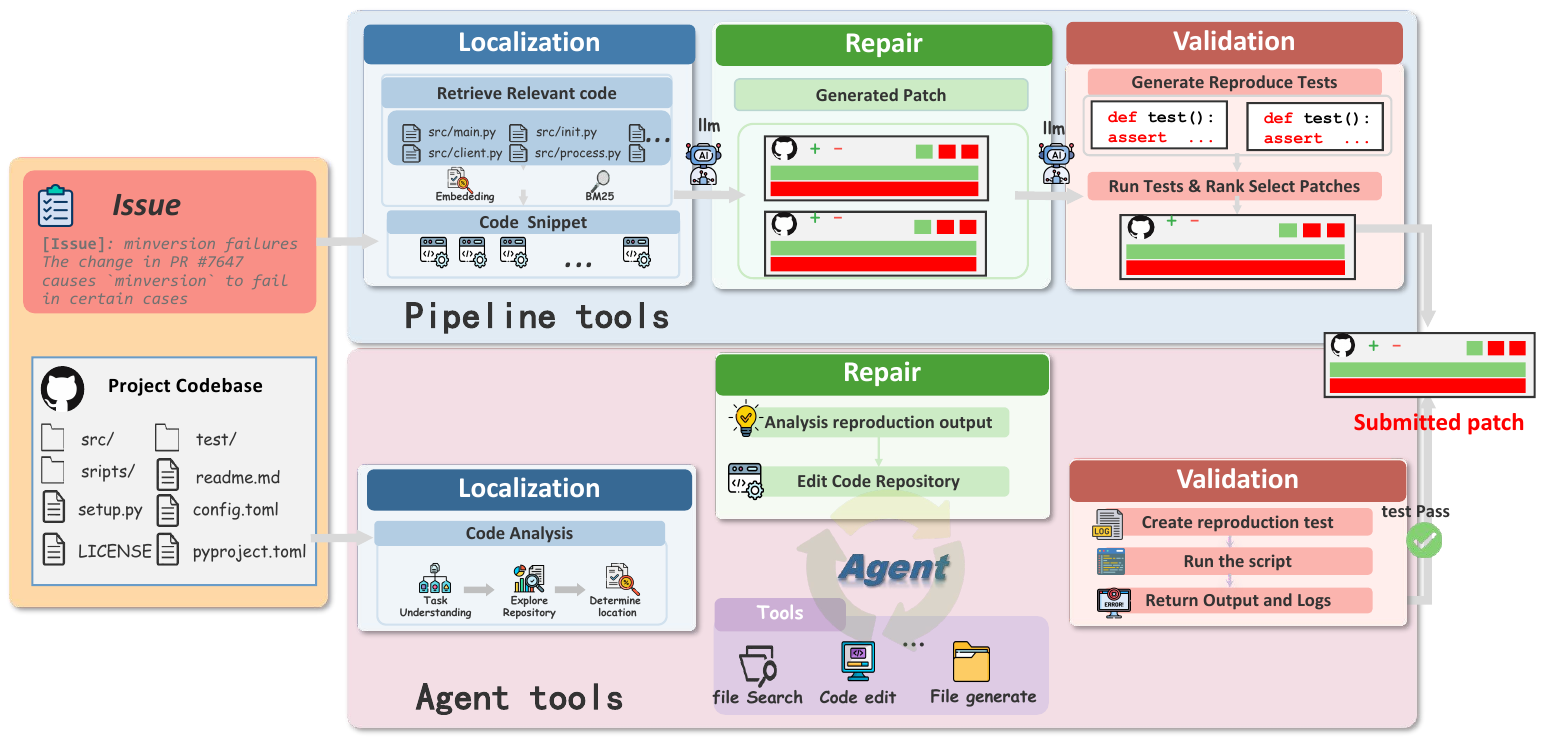}
    \caption{Workflow of the pipeline-based and agent-based tools.}
    \label{fig:overview}
    \vspace{-0.3cm}
\end{figure*}

LLM-based issue solving tools generally follow two distinct approaches: pipeline-based tools \cite{xia2024agentless,xie2025swe} and agent-based tools \cite{wang2024openhands, claude}, as shown in Figure \ref{fig:overview}. 
\textbf{Pipeline-based tools} decompose the task into fixed stages, such as localization, repair, and validation \cite{xia2025demystifying}, without dynamic decision-making or agentic behavior.  
Aider \cite{Aider} adds a preprocessing step where static code analysis builds a repository map
% (file/module summaries and dependency/declaration graphs) 
used to constrain retrieval, assemble prompts, and select context windows for the LLM. 
Agentless \cite{xia2024agentless} bypasses this step, using issue descriptions to retrieve relevant code elements directly. 
Regarding issue localization, SWE-Fixer \cite{xie2025swe} uses BM25 retrieval for issue localization, while Agentless combines LLMs with information retrieval methods.
Additionally, Repograph-enhanced Agentless improves code analysis by tracing dependencies and execution flow \cite{ouyang2024repograph}. Despite these improvements, these tools struggle with complex, multi-file issues due to their rigid workflows, limiting flexibility. 

In contrast, \textbf{agent-based tools} operate through iterative observation–action loops \cite{shinn2023reflexionlanguageagentsverbal,yao2023reactsynergizingreasoningacting}, enabling dynamic exploration and adaptive repair.
% \textbf{Agent-based tools} treat issue solving as an interactive process. 
Tools such as SWE-Agent \cite{yang2024swe}, OpenHands \cite{wang2024openhands}, and Anthropic's Tools+Claude adopt ReAct-style \cite{yao2023reactsynergizingreasoningacting} reasoning, interleaving natural language thought with tool usage. 
This design allows agents to iteratively navigate repositories, gather evidence, and refine candidate patches. These methods currently dominate the SWE-Bench-Verified leaderboard, demonstrating strong capability in handling complex issues that pipeline tools struggle with \cite{martinez2025dissecting}.
Building on this paradigm, several studies have proposed multi-agent extensions that split the repair workflow across specialized roles. For example, AgentScope \cite{gao2024agentscope} structures the process into sequential stages of reproducing, fixing, and testing, each handled by a dedicated agent with its own toolset. MarsCode Agent \cite{liu2024marscode} expands this idea by introducing a larger set of specialized agents and dynamically assembling them into static or adaptive pipelines, supported by a repository-level knowledge graph. While such designs improve modularity and can parallelize certain sub-tasks \cite{wu2023autogenenablingnextgenllm}, they remain fundamentally \textit{workflow-driven}: agents execute statically assigned roles in serial or parallel, passing results along a fixed pipeline \cite{zenml2025steerableblog}. This division of labor alleviates some bottlenecks but does not simulate the proactive, interactive collaboration \cite{wu2003study,herbsleb2003empirical} observed in human development teams, such as critiquing patches or correcting
misguided strategies.

\section{Methodology}

\subsection{Research Questions}

We formulate four research questions to uncover the current limitations of automated issue resolution tools and to inspire future directions for developing more reliable and robust solutions.

\begin{itemize}[leftmargin=*]
    \item \textbf{RQ1: Performance and Efficiency Analysis (\S\ref{sec:rq1})}. 
    How do state-of-the-art tools perform in automated issue solving considering success rates, task characteristics, and interaction efficiency?

    \item \textbf{RQ2: A Taxonomy of Failure Modes (\S\ref{sec:rq2})}. What are the common failure modes of the studied tools in automated issue solving?

    \item \textbf{RQ3: Analysis of Failure Modes (\S\ref{sec:rq3})}. How are failure modes distributed within this taxonomy, and what architecture-specific vulnerabilities do they reveal, particularly in relation to task difficulties?
    
    \item \textbf{RQ4: Mitigation Exploration with a Collaborative Architecture (\S\ref{sec:rq4})}. To what extent can the identified failures be mitigated using our proposed collaborative Expert–Executor framework?
\end{itemize}

\subsection{Studied Tools and Benchmark}

\subsubsection{Studied Tools}
We select three representative tools that reflect the dominant architectural paradigms in issue solving, all of which were top performers on the SWE-Bench leaderboard at the time of our study: 
\ding{182} OpenHands-CodeAct-2.1 \cite{neubig2024-openhands-codeact-2.1:}: an agentic tool (Rank 3 on SWE-Bench Verified as of February 2, 2025 with Claude-3.5-Sonnet),
\ding{183} Tools + Claude 3.5 Sonnet \cite{Tools-Claude}: a tool integration framework (Rank 9), and
\ding{184} Agentless-1.5 \cite{xia2024agentless}: the pipeline-based method.
To isolate the impact of architectural design on the results, all three tools are evaluated under the same backbone LLM (Claude 3.5 Sonnet). 
This selection allows us to examine how different architectures affect performance while keeping the backbone model constant.

\subsubsection{Benchmark}

As mentioned before, we employ SWE-Bench-Verified as the foundation for our study. We utilized publicly available, official experimental results instead of re-running the tools ourselves. Specifically, \textit{the data for our analysis—including execution logs, generated patches, and pass/fail statuses for all three studied tools—was sourced directly from the official SWE-bench Experiments repository} \cite{swe-bench-experiments-repo}. The repository hosts official submissions and their execution artifacts in a standardized format.

For our quantitative performance analysis (RQ1), we analyzed the complete official results of the three studied tools on the full 500-issue SWE-Bench-Verified benchmark. This allowed us to establish a baseline understanding of their overall performance and limitations. In our qualitative failure analysis (RQ2 \& RQ3), we extract data regarding the solution status across the three selected tools, retaining cases where at least one tool failed to provide a solution, resulting in 329 records.
Since manual analysis of all these cases would be time-consuming, we randomly sampled 150 issues from this set with a 95\% confidence level and a margin of error of 5\% \cite{wikipedia_sample_size,baltes2021samplingsoftwareengineeringresearch,israel1992determining}.
This ensures sufficient representativeness while maintaining analytical feasibility. The sample exhibits a realistic difficulty distribution with 92 medium issues (61.3\%), 38 easy issues (25.3\%), and 20 hard issues (13.3\%), reflecting the complexity spectrum of real-world software engineering tasks.

\section{Performance and Efficiency Analysis (RQ1)}\label{sec:rq1}

\subsection{Overall Performance Comparison (RQ1.1)}\label{sec:rq1-1}
\begin{wrapfigure}{r}{0.35\textwidth}
    \vspace{-0.5cm}
    \setlength{\abovecaptionskip}{-0.2cm}
    \centering
    \includegraphics[width=\linewidth]{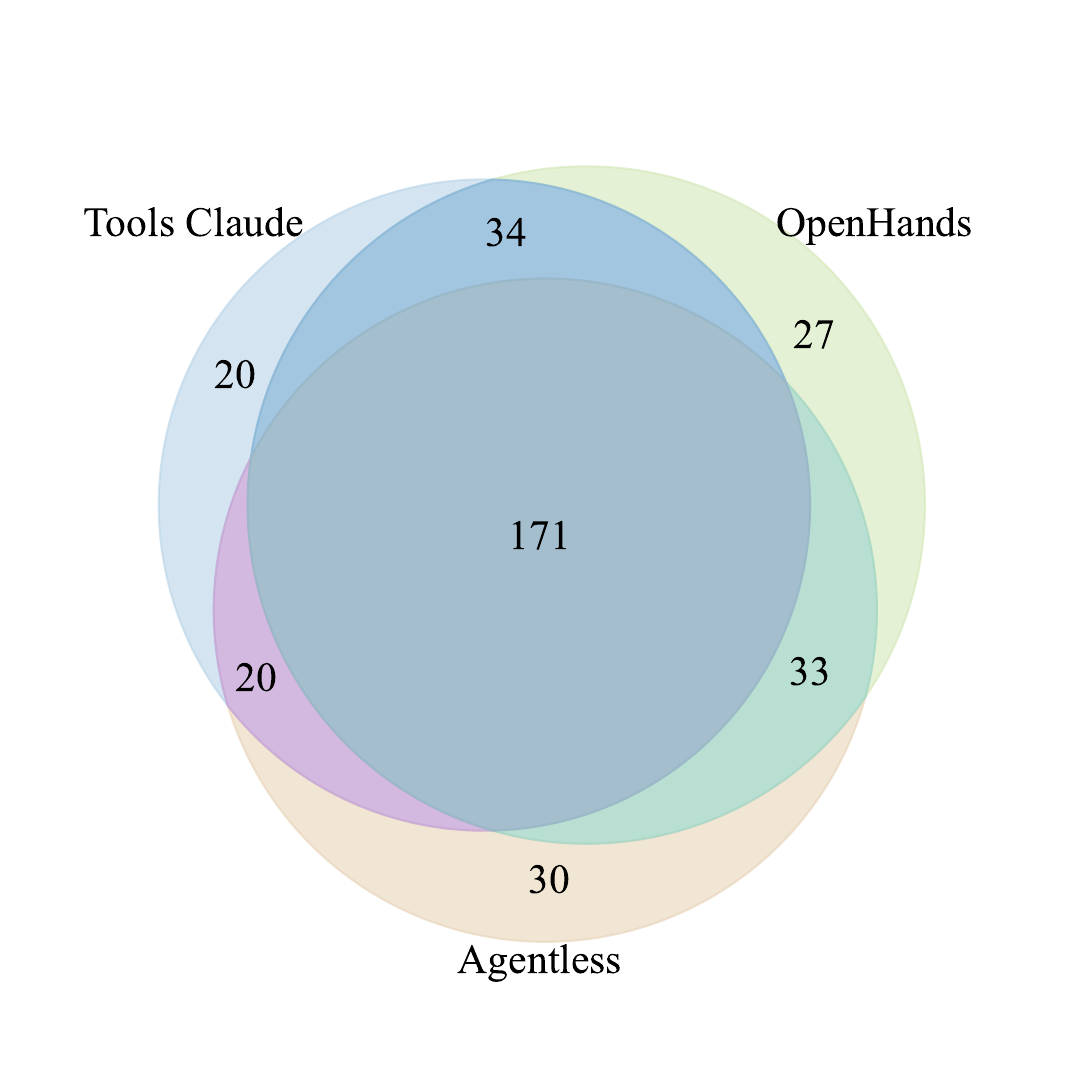}
    \caption{Venn diagram of resolved cases of the studied tools.}
    \label{fig:rq1_overlap}
    \vspace{-10pt}
\end{wrapfigure}
We evaluated three representative tools on 500 issues from SWE-bench Verified: \textbf{OpenHands CodeAct v2.1} (interactive agent with optimized scaffolding), \textbf{Agentless} (pipeline-style approach), and \textbf{Tools Claude} (interactive agent with standard scaffolding).
Across the full set, OpenHands solved 53.0\% (265/500), Agentless 50.8\% (254/500), and Tools Claude 49.0\% (245/500). The absolute differences are modest (within 4\%), indicating broadly comparable effectiveness on the SWE-bench Verified benchmark. Notably, OpenHands and Tools Claude share the same base model and a similar toolset, yet still differ by 4.0 points. While we do not attribute this gap to any single factor, practical implementation details (\textit{e.g.}, how file navigation and tool usage are instructed, or workspace path conventions) appear to matter in aggregate.

Beyond overall rates, we also presented the issue-solving distribution diagrams of these tools, as shown in Figure~\ref{fig:rq1_overlap}.
The results suggest complementary strengths across different tools. Specifically, OpenHands and Tools Claude solved 205 issues in common, but each also solved a distinct subset (OpenHands: +60; Tools Claude: +40), and Agentless contributed 30 unique solutions. This indicates that, despite similar aggregate accuracy, the tools excel in solving different instances, suggesting potential benefits for ensemble or portfolio-style usage.

\vspace{1mm}
\begin{custommdframed}
\textbf{Finding 1:} 
The three studied tools achieve similar overall issue solving rates (49--53\%) but exhibit complementary coverage at instance level, suggesting the potential benefits of exploring ensemble methods or selection strategies to optimize issue resolution.
\end{custommdframed}
\vspace{1mm}

\subsection{Task Characteristics and Performance Correlation (RQ1.2)}\label{sec:rq1-2}
% To understand the fundamental drivers of agent performance,
We further examined how intrinsic issue characteristics correlate with solving performance, focusing on \ding{182} issue description length, \ding{183} task difficulty (measured by estimated human solving time), and \ding{184} task complexity (measured by the number of files modified in the ground-truth human patch, \textit{i.e.}, single-file vs. multi-file).

\begin{wrapfigure}{r}{0.5\textwidth}
    \vspace{-10pt}
    \setlength{\abovecaptionskip}{0.1cm}
    \centering
    \includegraphics[width=0.9\linewidth]{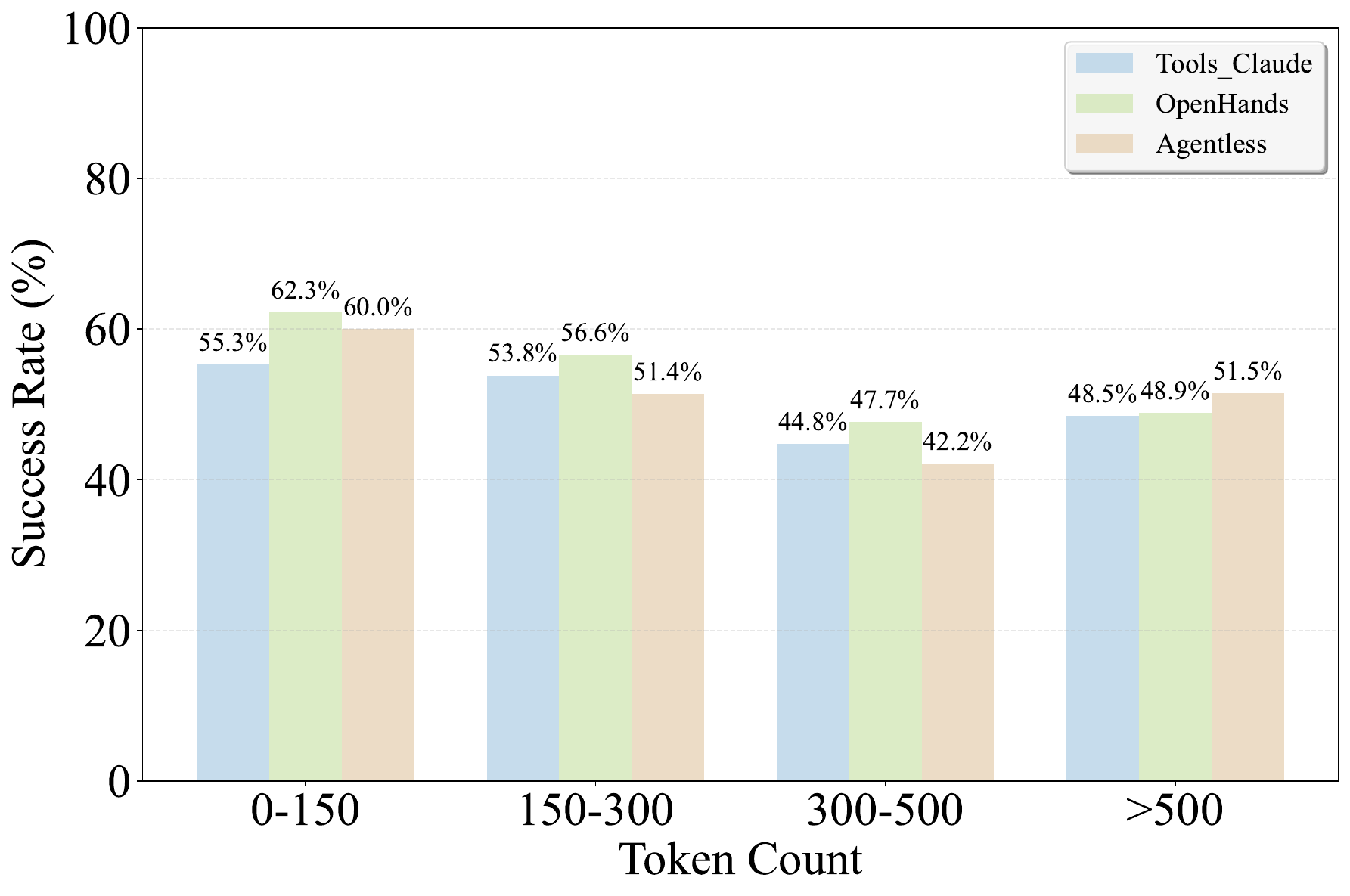}
    \caption{Impact of issue description length.}
    \label{fig:token_count_success_rates}
    \vspace{-10pt}
\end{wrapfigure}

\textbf{Issue Description Length.} To ensure a robust analysis, we binned the issues into four quartiles based on description token count, creating groups of nearly equal size. 
The relationship between issue description length and success rate is not consistent, but rather reflects a trade-off between problem complexity and contextual clarity. 
As shown in Figure~\ref{fig:token_count_success_rates}, the highest performance is achieved on the shortest issues (0-150 tokens), likely because brevity often correlates with simpler, more localized problems that are easier to resolve. However, as descriptions lengthen towards the 500-token mark, success rates decline, suggesting that this range often represents an increase in intrinsic problem complexity without providing compensatory detail. This trend then reverses for the longest issues (\textgreater 500 tokens), where length is typically a function of rich, prescriptive context—such as detailed error logs, code snippets, and explicit steps—that provides a clear path to a solution, outweighing the task's inherent difficulty.

\begin{figure}[t]
    \centering
    \setlength{\abovecaptionskip}{0.1cm}
    \begin{subfigure}[t]{0.48\textwidth}
        \centering
        \includegraphics[width=\linewidth]{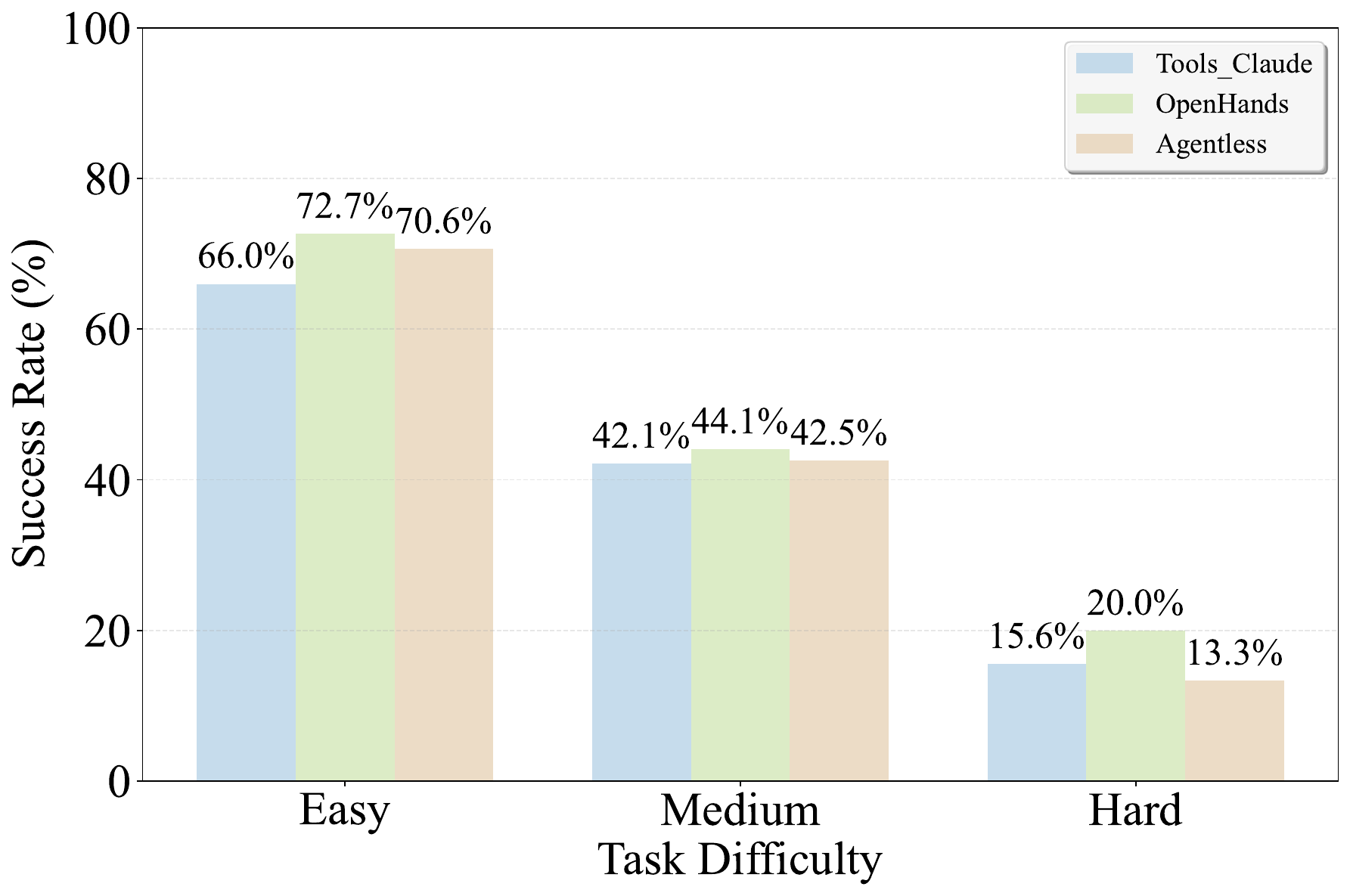}
        \caption{Issue solving rates across different difficulty levels.}
        \label{fig:difficulty_success_rates}
    \end{subfigure}\hfill
    \begin{subfigure}[t]{0.48\textwidth}
        \centering
        \includegraphics[width=\linewidth]{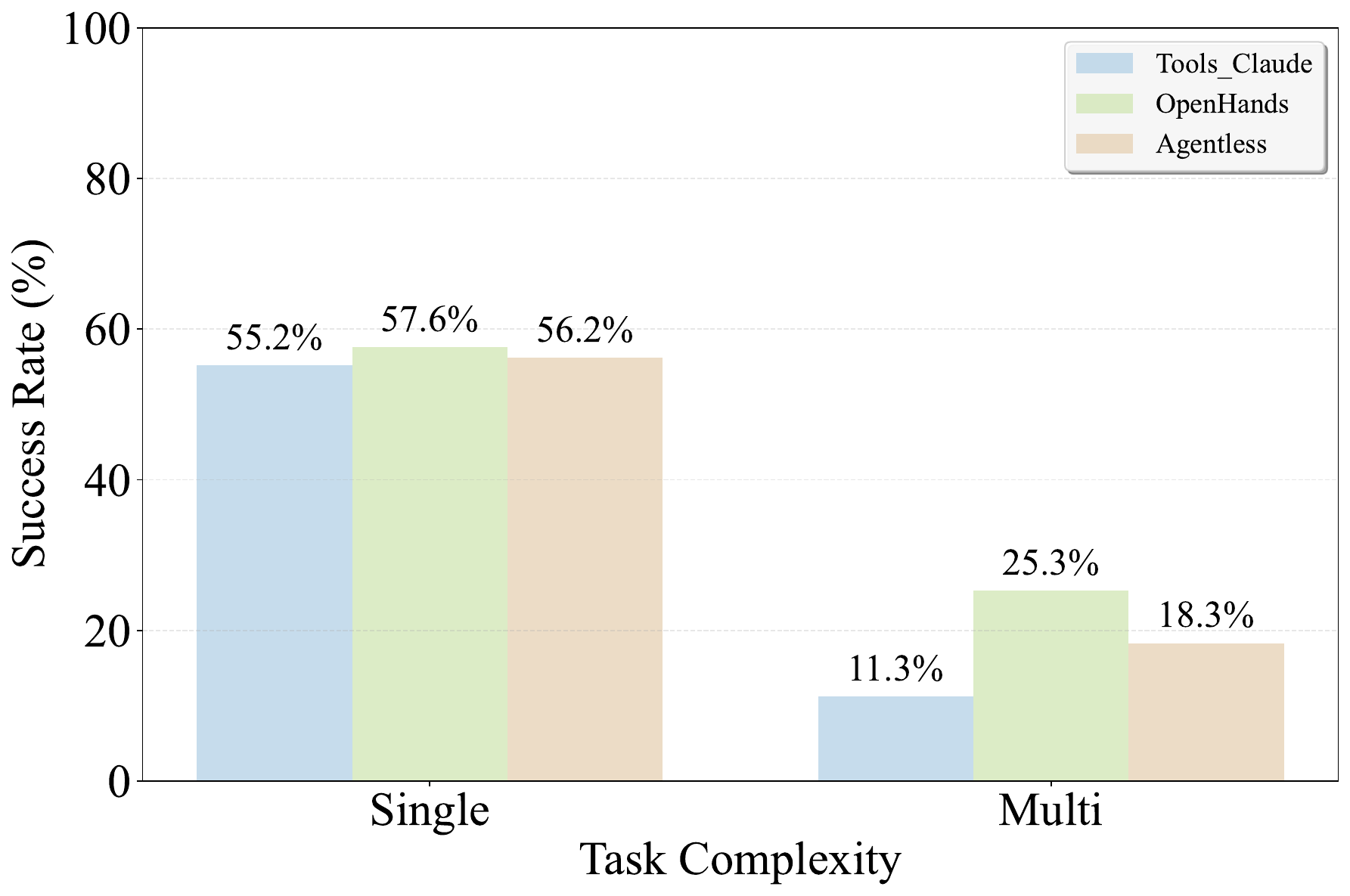}
        \caption{Issue solving rates across task complexity.}
        \label{fig:file_complexity_success_rates}
    \end{subfigure}
    \caption{Impact of task characteristics on issue solving performance.}
    \label{fig:rq1_complexity_panels}
    \vspace{-0.3cm}
\end{figure}

\textbf{Task Difficulty.} As shown in Fig.~\ref{fig:difficulty_success_rates}, the performance of LLM-based agents decreases consistently as the task difficulty increases from Easy to Medium and Hard. For example, OpenHands drops to 44\% on Medium tasks, while Agentless and Tools Claude experience declines of more than 20 percentage points. On Hard tasks (\(\geq 1\,\text{h}\)), all tools converge to a low success range of 13–20\%. 
Across all three tools, success rates show a strong negative correlation with human-perceived difficulty, indicating that the benchmark’s challenges align well with real-world developer effort. These tools excel at solving well-defined problems that a human developer can resolve in under 15 minutes. However, they face significant challenges with more complex, multifaceted problems that require deep reasoning, strategic planning, and sustained problem-solving over extended periods.

\textbf{\textit{Task Complexity.}}
In this section, we investigate how issue complexity, measured by the number of files modified in the ground-truth human patch, affects tool performance.
Specifically, ``single-file'' vs. ``multi-file'' refers to the number of files modified in the ground-truth human patch for an issue.
% (\textit{i.e.}, how many files human developers changed to resolve it).
As shown in Figure~\ref{fig:file_complexity_success_rates}, \textit{Single-file} modifications are handled reasonably well by all tools (55-58\% success). However, \textit{multi-file} coordination exposes substantial differences: OpenHands maintains a success rate of 25.3\%, whereas Tools Claude (Interactive Agent with Standard Scaffolding) drops to 11.3\% and Agentless to 18.3\%. These sharper declines highlight that cross-file coordination is a key bottleneck.
Agentless, as a pipeline architecture, appears less adept at producing multi-file patches when the human ground truth involves changes across multiple files. Among 497 successful generated Agentless patches, only 6 touched more than one file. For 13 issues whose human fixes required multiple files, Agentless still passed tests with single-file edits in 12 cases. This pattern suggests that \textit{the space of valid fixes is not unique and a patch that perturbs fewer locations than the human edit can still make the test suite pass}, although such minimal edits may trade off generality or long-term maintainability.

\vspace{1mm}
\begin{custommdframed}
\textbf{Finding 2:} 
Issue solving rates consistently degrade as task complexity increases, particularly when handling multi-file coordination. OpenHands demonstrates the greatest resilience, maintaining performance even in more complex scenarios. In contrast, Agentless and Tools Claude experience more significant drops, underscoring the need for architectures specifically designed to handle intricate, multi-step problems and cross-file dependencies.
\end{custommdframed}
\vspace{1mm}

\subsection{Interaction Efficiency (RQ1.3)}\label{sec:rq1-3}

\begin{figure}[t]
    \centering
    \setlength{\abovecaptionskip}{0.1cm}
   \begin{subfigure}[t]{0.45\textwidth}
        \centering
        \includegraphics[width=1\linewidth]{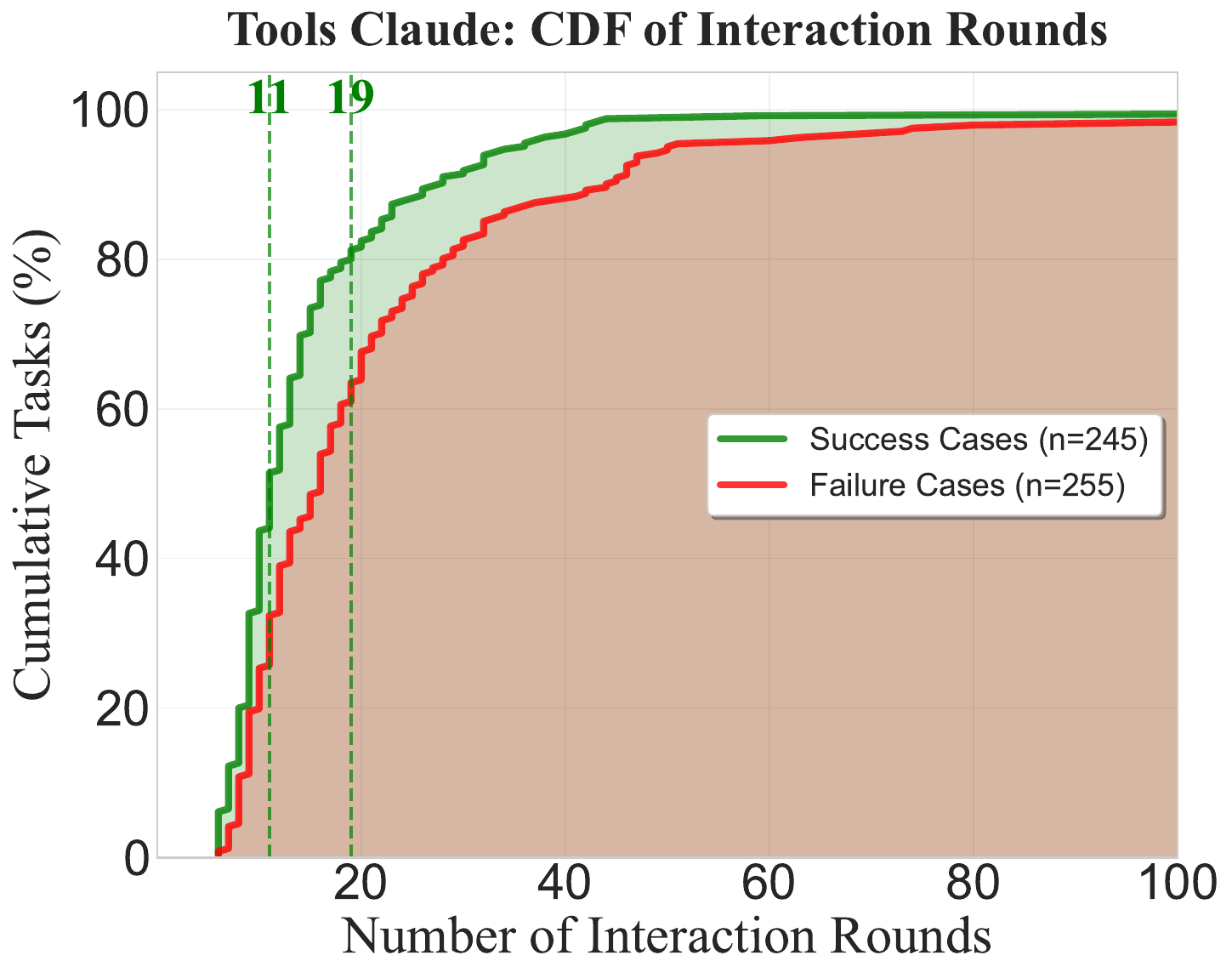}
        \caption{Interaction rounds for Tools Claude.}
        \label{fig:tools_claude_cdf}
    \end{subfigure}\hfill
      \begin{subfigure}[t]{0.45\textwidth}
        \centering
        \includegraphics[width=1\linewidth]{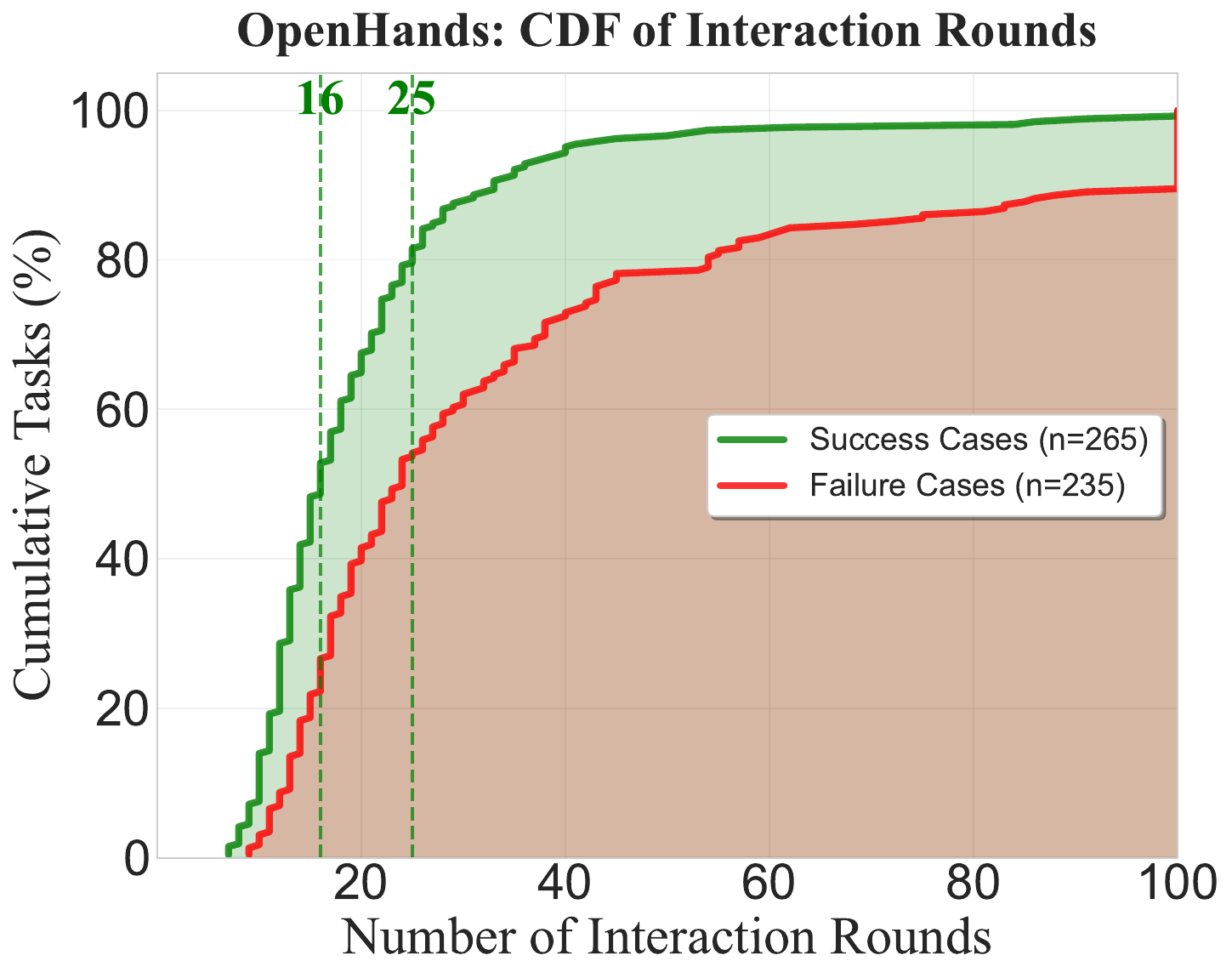}
        \caption{Interaction rounds for OpenHands.}
        \label{fig:openhands_cdf}
  \end{subfigure}
  \caption{Interaction efficiency comparison between successful and failed cases.}
  \vspace{-0.3cm}
\end{figure}

To analyze interaction efficiency, we examine the cumulative distribution function (CDF) of interaction rounds for successful and failed tasks in two agent-based tools: OpenHands and Tools Claude (Figure~\ref{fig:tools_claude_cdf} and Figure~\ref{fig:openhands_cdf}). As seen from the results, both tools show a clear separation between successful and failed tasks. Tools Claude resolves 80\% of its successful tasks within a brisk 19 rounds (median 11), while OpenHands requires 25 rounds for the same success rate (median 16). Crucially, failed tasks exhibit a long-tail effect \cite{dean2013tail}. OpenHands, for instance, requires 54 rounds to cover 80\% of its failures, suggesting it often gets stuck in prolonged, unproductive explorations when a solution is not readily found. The analysis reveals that a majority of successful resolutions occur within approximately 25 interaction rounds, highlighting a critical window for efficient problem-solving.

% \noindent \textbf{Implications for Agent Design.}
%The analysis shows that tasks should ideally resolve within 25 rounds to be successful. Tools Claude's strategy offers quick solutions but risks failure in large tasks, while OpenHands provides a more robust solution at the expense of efficiency. Future agents could benefit from hybrid strategies that adapt to task complexity. 
% Additionally, an optimal timeout policy based on diminishing returns (around 35 rounds) can help conserve resources.

\vspace{1mm}
\begin{custommdframed}
\textbf{Finding 3:} 
% Agent efficiency is determined by interaction strategies. 
The value of continued interaction in agent-based problem solving exhibits sharply diminishing returns. Most successes are found within the first 25 rounds. Moreover, agents are far more likely to enter a long-tail failure pattern, engaging in extended but futile explorations. This highlights a critical need for tools that can dynamically assess progress and halt unproductive attempts.
\end{custommdframed}
\vspace{1mm}

\section{Failure Modes in LLM-based Automated Issue Solving (RQ2)}\label{sec:rq2}
Our performance analysis in RQ1 revealed that even state-of-the-art agents fail on a substantial portion of tasks. 
More importantly, the efficiency analysis highlight systemic behavioral patterns, such as agents getting trapped in prolonged, unproductive loops, but it does not explain the underlying causes of these failures. 
To move beyond \textit{what} happens to \textit{why} it happens, in this RQ, we perform a qualitative analysis to build a structured classification of failure modes tailored to the issue-solving lifecycle. To achieve this, we conduct a systematic manual analysis procedure to identify and categorize recurrent failure patterns from automated issue solving tools' raw execution traces. 

\subsection{Procedure for Taxonomy Construction}\label{sec:manual_analysis}

To construct our taxonomy of failure modes (RQ2) and to characterize failure patterns (RQ3), we designed a \textbf{multi-stage manual analysis procedure} that combined open coding \cite{wicks2017coding}, iterative taxonomy refinement, and rigorous annotation protocols. Our process followed established practices in qualitative software engineering research, while being tailored to the sequential execution traces of LLM-based code agents.

\paragraph{Stage 1: Pilot Analysis and Initial Taxonomy Construction.}
We first conducted a \emph{pilot analysis} to establish an initial taxonomy of failure modes. For this purpose, we selected 50 execution traces randomly from \texttt{OpenHands-CodeAct-2.1}, a representative example of agent tool in our study. Its reasoning traces provided a broad spectrum of failure phenomena, making it suitable for seeding the taxonomy.
Two annotators (both Ph.D. students with more than seven years of Python development experience and prior research in automated program repair) independently analyzed sampled failed cases. Following the principles of open coding, each annotator proposed tentative categories by grouping recurrent failure manifestations, based on both tool execution logs and associated test outcomes. After independent analysis, the annotators met to consolidate their findings, merging overlapping categories and clarifying ambiguous definitions. 

This stage took approximately 100 person-hours and yielded an \emph{initial codebook}\footnote{The final codebook can be found in our replication package.} comprising nine preliminary categories, which served as the foundation for large-scale annotation in Stage 2. 

\paragraph{Stage 2: Full Annotation and Taxonomy Refinement.}
In the second stage, we expanded the annotation to a sample of 150 failed cases drawn evenly across all three studied tools (OpenHands-CodeAct-2.1, Tools + Claude 3.5 Sonnet, and Agentless-1.5). Annotation was conducted by \textbf{four annotators}, including the two from Stage 1 and two additional researchers (both with over five years of software engineering research and industry programming experience). 
To facilitate annotation, we pre-processed all execution traces into a structured format and developed a lightweight \textbf{web-based annotation platform}. The platform presented annotators with each interaction round of the agent, including its \emph{thought process}, \emph{executed action}, and the corresponding \emph{tool feedback}. Each case was independently annotated by two annotators in parallel, ensuring every instance received at least dual coding. Annotators were instructed to:
\begin{enumerate}[leftmargin=*]
    \item Assign one or more failure modes from the codebook.
    \item Identify the specific step(s) in the execution trace where the critical failure emerged.
      \item Document the underlying cause(s) and observable effect(s).
\end{enumerate}

Annotators could flag any instance not covered by the current codebook and submit a label proposal within the platform (additions, deletion or definition edits). Upon submission, the platform automatically notified all annotators; proposals were available for immediate confirmation by any annotator and were reviewed by the four annotators on a rolling basis. Disagreements were resolved via discussion, with final arbitration provided by a senior author with more than ten years of APR research experience.
After coding the full set of 150 cases, no additional categories were required beyond the refined Stage-1 codebook, indicating \emph{theoretical saturation} for our sample. Stage 2 required approximately 620 person-hours of effort.

To assess annotation reliability, we computed \textbf{Cohen's Kappa} scores across all four annotators. Agreement was measured on two dimensions: (i) assigned failure modes; (ii) the step index in the agent's interaction trajectory where the decisive failure emerged (interactive tools only; not applicable to Agentless). Pairwise Kappa \cite{P_rez_2020} values ranged from \textbf{0.72 to 0.77} across dimensions; for (ii), \textbf{Kappa} was computed on the interactive subset, exceeding the conventional 0.6 threshold for substantial agreement \cite{article,li2023kappa}. Annotation disagreements were first reconciled by the original annotator pair. The small remainder that persisted after pairwise reconciliation (less than 8\% of cases) were resolved in a group discussion with final adjudication by the senior author, yielding a fully adjudicated gold label for every case. 

\subsection{Taxonomy of Failure Modes}\label{sec:taxonomy}

Finally, we identified a total of 342 failure instances across 150 issues. These were organized into three phases—Location, Repair, and Iteration \& Validation—encompassing 9 primary failure categories and 25 fine-grained subcategories.

\begin{wrapfigure}{r}{0.5\textwidth}
    \vspace{-10pt}
    \centering
    \setlength{\abovecaptionskip}{0.1cm}
    \includegraphics[width=\linewidth]{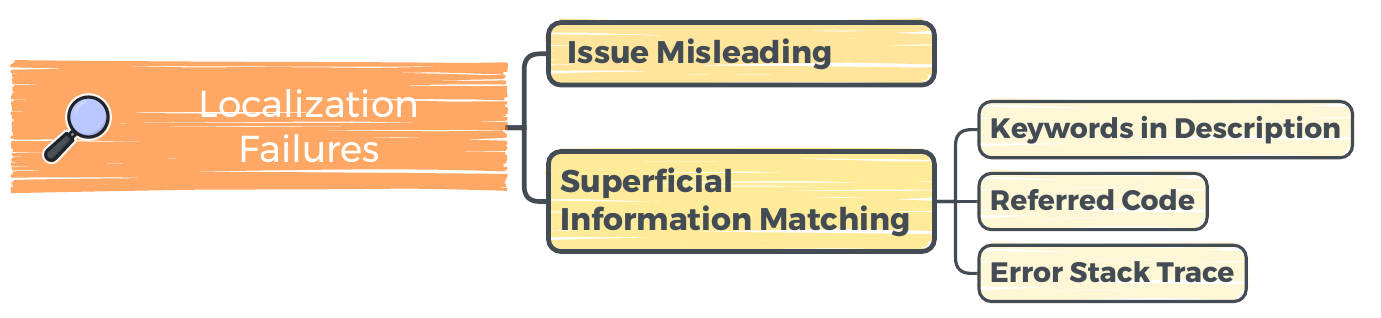}
    \caption{Failure modes in Localization stage.}
    \label{fig:mode13}
    \vspace{-10pt}
\end{wrapfigure}

\subsubsection*{\textbf{A. Localization Failures}}
Failures in this initial stage arise when the tool misinterprets the issue or fails to pinpoint the correct code regions requiring modification. Such errors are often catastrophic, as they misdirect the subsequent repair and validation steps. Figure~\ref{fig:mode13} illustrates a taxonomy of such failure modes. Failures in this stage
can be further divided into the following sub-
categories:

\begin{wrapfigure}{r}{0.72\textwidth} 
\vspace{-0.4cm}
\begin{tcolorbox}[top=0.1cm,bottom=0.1cm,
  colback=white,colframe=black!50,boxrule=0.5pt,arc=0pt]
\textbf{\texttt{Issue Misleading Example}} % 加粗并打字机字体

\begin{lstlisting}[basicstyle=\scriptsize\ttfamily,
                   breaklines=true,
                   escapeinside={(*@}{@*)}]
This would mean one changes the '(*@\textcolor{red}{\_SetOutputMixin}@*)' to add: 
* another argument `dtypes` to `_wrap_in_pandas_container`. 
\end{lstlisting}

\end{tcolorbox}
\vspace{-0.4cm}
\end{wrapfigure}

\noindent\textbf{\textit{A1: Issue Misleading}.} 
This failure arises when the tool follows misleading implementation suggestions in the issue description rather than independently diagnosing the underlying problem, leading it to modify incorrect locations in the codebase.
As shown in the above example (\texttt{scikit-learn\_\_scikit-learn-25102}), the reporter not only described the problem but also proposed a detailed—yet incorrect—solution. Misled by this guidance, the tool focused on output-wrapping logic in \texttt{\_set\_output.py}, while the true cause lay earlier in the pipeline, where inputs were prematurely cast to NumPy arrays in \texttt{sklearn/base.py} and \texttt{sklearn/feature\_selection/\_base.py}. By uncritically following the suggested roadmap, the agent treated symptoms rather than the root fault.

\noindent\textbf{\textit{A2: Superficial Information Matching}.} 
This failure arises when the tool lacks a deep semantic understanding of the codebase and instead depends on shallow heuristics for localization. It can be further divided into the following sub-categories:

\noindent\textit{A2.1: Keywords in Description.} The tools may rely on naive keyword searches derived from issue descriptions, which can misdirect them to functionally related but incorrect files.
% This occurs when the tool's internal thought identifies a keyword for broad search commands (\textit{e.g.}, \texttt{grep}).
For example, in \texttt{pylint-dev\_\_pylint-6386}, the description mentioned a buggy ``verbose'' short option. The tool performed multiple \texttt{grep ``verbose''} commands, returning many candidates (documentation, config files, checker logic). It incorrectly modified \texttt{options.py} containing option metadata, while the actual bug spanned another three files handling command-line argument processing: \texttt{pylint/config/argument.py}, \texttt{arguments\_manager.py}, and \texttt{utils.py}. 

\noindent\textit{A2.2: Referred Code.} 
Issue descriptions often reference specific code snippets as examples, and tools may use the provided example code to identify and modify the corresponding module, even if the true fault lies in a different, interacting module.
For instance, in \texttt{django\_\_django-14034}, the reporter showed that \texttt{MultiValueField} failed to enforce required subfields. The agent focused narrowly on this class and patched its validation logic. However, the actual fix required modifying widget attribute handling in \texttt{boundfield.py}. The tool never explored this location because its investigation was anchored solely to the class mentioned in the example.

\noindent\textit{A2.3: Error Stack Trace.}
Tools may also be misled by stack traces \cite{inproceedings} from the issue description or their own reproduction scripts, leading them to attempt localized patches without examining the broader call chain to identify the root cause. For example, in \texttt{sympy\_\_sympy-12420}, the tool observed a crash culminating in the trace below, and it directly modifies the \texttt{radsimp.py} file to handle this \texttt{IndexError}. However, the root cause was in an upstream file, \texttt{sqrtdenest.py}, which was passing an empty tuple to \texttt{split\_surds}. A proper fix required adding a precondition check in the calling function, not handling the crash downstream.

\begin{wrapfigure}{r}{0.5\textwidth}
\vspace{-0.4cm}
\begin{tcolorbox}[top=0.1cm,bottom=0.1cm,
                  colback=white,colframe=black!50,
                  boxrule=0.5pt,arc=0pt,width=0.5\textwidth]
\textbf{\texttt{Error Stack Trace Example}}

\begin{lstlisting}[basicstyle=\scriptsize\ttfamily,
                   breaklines=true,
                   escapeinside={(*@}{@*)}]
File "(*@\textcolor{red}{sympy/simplify/radsimp.py}@*)", line 1068, in _split_gcd
    g = a[0]
IndexError: tuple index out of range
\end{lstlisting}
\end{tcolorbox}
\vspace{-0.4cm}
\end{wrapfigure}

\subsubsection*{\textbf{B. Repair Failures}}

Once a location is identified, the tools generate a code patch to repair the faults and Figure~\ref{fig:mode2} illustrates a taxonomy of such failure modes. Failures in this stage can be further divided into the following sub-categories:

\begin{wrapfigure}{r}{0.65\textwidth}
    \vspace{-10pt}
    \centering
    \setlength{\abovecaptionskip}{0.1cm}
    \includegraphics[width=\linewidth]{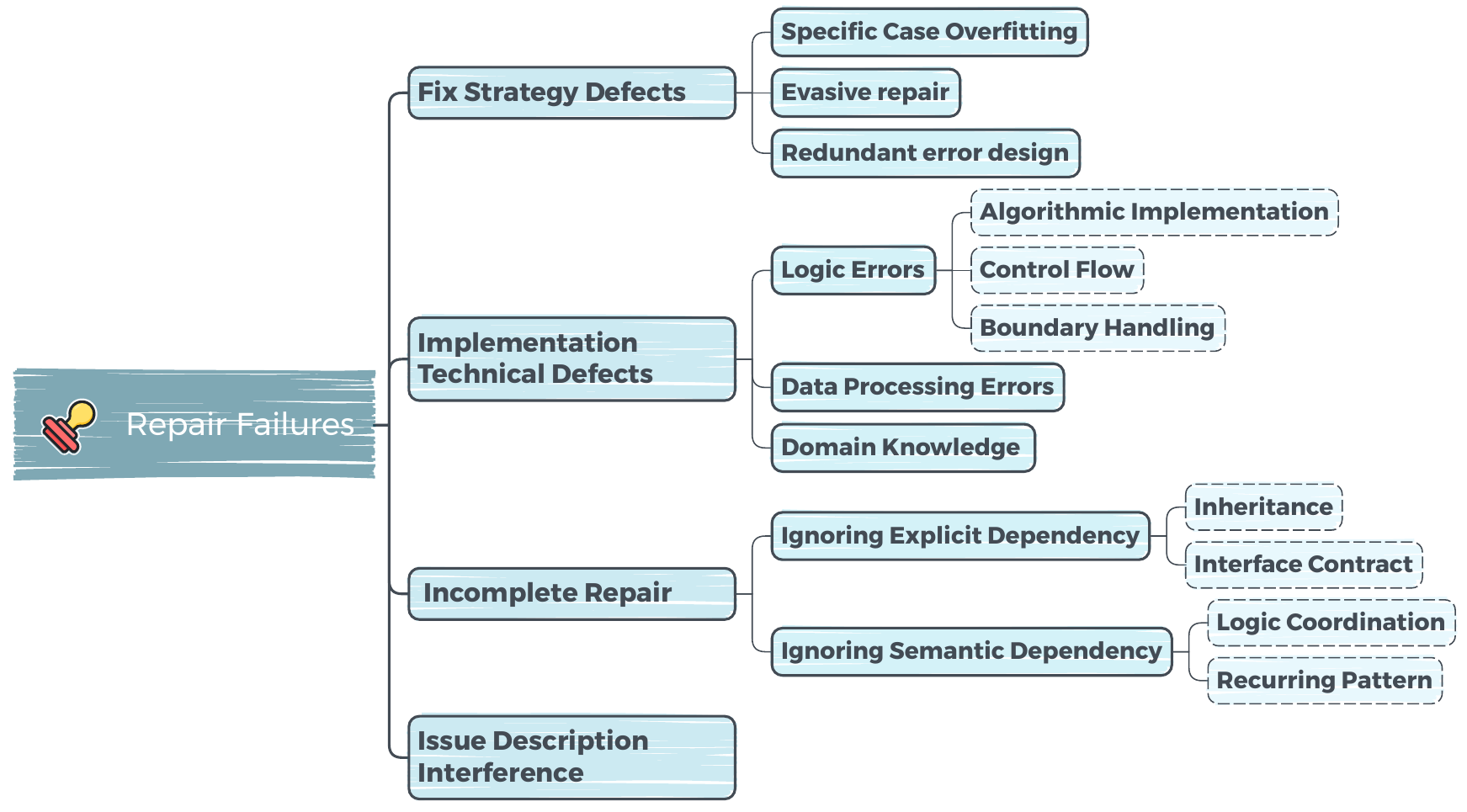}
    \caption{Failure modes in Repair stage.
    % Problem-level overlap analysis showing complementary strengths across tools.
    }
    \label{fig:mode2}
    \vspace{-10pt}
\end{wrapfigure}

\noindent\textbf{\textit{B1: Fix Strategy Defects.}} 
This category of failure arises from flaws in the tool's high-level repair strategy, preventing the tool from effectively resolving the issue.

\noindent\textit{B1.1: Specific Case Overfitting.} The generated patch may be narrowly tailored to the specific scenario presented in the issue description, leading to brittle and incomplete patches \cite{10.1145/2786805.2786825}.
In the example below (\texttt{django\_\_django-13551}), which reported that password reset tokens were not invalidated when a user's email changed, the tool implemented a solution that directly references the \texttt{email} attribute. This solution overfits by assuming the user model always has an attribute named \texttt{`email'}. The correct, generalizable solution was to use \texttt{user.get\_email\_field\_name()} to dynamically determine the email field. 

\begin{tcolorbox}[top=0cm,bottom=0cm,
                  colback=white,colframe=black!50,
                  boxrule=0.5pt,arc=0pt]

\textbf{\texttt{Specific Case Overfitting Example}}

\begin{lstlisting}[language=diff,
                   basicstyle=\scriptsize\ttfamily,
                   breaklines=true,
                   escapeinside={(*@}{@*)}]
- return str(user.pk) + user.password + str(login_timestamp) + str(timestamp)
+ (*@\fbox{email}@*) = (getattr(user, '(*@\fbox{email}@*)', '') or '').lower().strip()
+ return str(user.pk) + user.password + (*@\fbox{email}@*) + str(timestamp)
\end{lstlisting}

\end{tcolorbox}

\noindent\textit{B1.2: Evasive Repair.}
This failure arises when the tool addresses a bug's symptom by wrapping problematic code in defensive constructs like broad \texttt{try-except} blocks or null checks, rather than fixing the underlying logical flaw. This prevents crashes but masks root causes.

As shown in \texttt{django\_\_django-16938}, a \texttt{FieldError} arises during serialization when a ManyToMany custom manager with \texttt{select\_related} conflicted with the serializer's \texttt{.only("pk")} optimization. The tool's evasive fix used \texttt{except: pass}, skipping the optimization instead of resolving the conflict. The correct solution required properly chaining \texttt{select\_related()} and \texttt{only("pk")}.

\begin{wrapfigure}{r}{0.35\textwidth} % r=右侧, l=左侧, 宽度=0.45倍行宽
\vspace{-0.3cm}
\begin{tcolorbox}[top=0.1cm,bottom=0.1cm,
                  colback=white,colframe=black!50,
                  boxrule=0.5pt,arc=0pt,width=0.35\textwidth]

\textbf{\texttt{Evasive Repair Example}}

\begin{lstlisting}[language=diff,
                   basicstyle=\scriptsize\ttfamily,
                   breaklines=true,
                   escapeinside={(*@}{@*)}]
+  qs = getattr(obj, field.name)
+  (*@\textcolor{red}{try}@*):
+     qs = qs.only("pk")
+  (*@\textcolor{red}{except}@*):
+     (*@\textcolor{red}{pass}@*)
+  return qs.iterator()
\end{lstlisting}

\end{tcolorbox}
\vspace{-0.3cm}
\end{wrapfigure}

\noindent\textit{B1.3: Redundant Erroneous Implementation.} 
The tools may introduce overly complex solutions by re-implementing existing logic or ignoring tool extension points, adding redundant code and potentially new bugs \cite{yang2024demystifying}.
In \texttt{django\_\_django-13449}, the agent ignored Django's standard \texttt{as\_sqlite} method and re-implemented the logic from scratch. The resulting solution was not only redundant but also ultimately flawed.

\noindent\textbf{\textit{B2: Implementation Details Defects.}}
This failure occurs when the tool's overall repair strategy is sound, but the implementation contains technical errors, leading to incorrect or failing patches.

\noindent\textit{B2.1: Logic Errors.}
This group covers fundamental flaws in the patch's reasoning and execution flow. We identify three distinct sub-categories of such failures: \textit{algorithmic error}, \textit{flawed control flow}, and \textit{inadequate boundary handling}. 
An example of an algorithmic error appears in \texttt{sympy\_\_sympy-23413}, where the agent mistakenly conflated two separate mathematical operations. For flawed control flow, in \texttt{matplotlib\_\_matplotlib-22871}, the agent oversimplified a conditional statement, failing to preserve the nuanced logic required to handle other cases correctly. Finally, inadequate boundary handling is illustrated by \texttt{django\_\_django-11141}, where removing a restrictive check introduced a new bug by causing empty directories—an edge case—to be misclassified.

\noindent\textit{B2.2: Data Processing Errors.}  
This category refers to mistakes in manipulating or transforming program data. Typical cases include incorrect type casting, variable scope mismanagement (\textit{e.g.}, using a local variable as if it were global) \cite{ouyang2025empirical}, or improper data formatting. Such errors often propagate downstream, resulting in crashes or incorrect behavior.  
 
\noindent\textit{B2.3: Insufficient Domain Knowledge.}  
This category arises when the tool lacks knowledge of external tools, protocols, or library-specific conventions \cite{zhang2025understandingmitigatingerrorsllmgenerated} that are unknown to the agent and not evident from the local code. Correct fixes may depend on such conventions or established practices. For example, in \texttt{django\_\_django-11239}, the task was to forward SSL parameters from Django's \texttt{dbshell} to PostgreSQL's \texttt{psql}. The tool extracted the parameters but incorrectly passed them as command-line arguments, whereas \texttt{psql} requires environment variables (e.g., \texttt{PGSSLMODE}, \texttt{PGSSLCERT}). This reflects insufficient domain knowledge of PostgreSQL's client interface.

\noindent\textbf{\textit{B3: Incomplete Repair.}}
This failure occurs when the tool addresses only part of the problem, leaving other necessary changes unimplemented.

\noindent\textit{B3.1: Ignoring Explicit Dependencies.} This happens when the fix may overlook structural dependencies in the code. For example, the tool may modify a child class without updating the parent class, or adjust one implementer of an interface while neglecting others. Such oversights violate explicit structural contracts of the codebase and can introduce inconsistencies or new bugs.

\noindent\textit{B3.2: Ignoring Semantic Dependencies.} The tool may also fail to account for logical dependencies that are not explicitly enforced by the code structure, resulting in partial fixes that leave related parts of the tool broken or out of sync. This includes cases where multiple components must remain consistent or recurring code patterns require simultaneous updates.

\noindent\textbf{\textit{B4: Issue Interference.}}
This failure arises when the tool follows misleading or overly prescriptive issue descriptions instead of validating intended behavior, leading to flawed patches. For example, in \texttt{sympy\_\_sympy-15599}, it adopted the report's suggestion to alter modular simplification, which fixed \texttt{Mod(3*i,2)} but broke other cases. The correct fix required handling integer coefficients relative to the modulus, beyond the issue's guidance.

\subsubsection*{\textbf{C. Iterative Verification Failures.}}
\begin{wrapfigure}{r}{0.65\textwidth}
    \vspace{-10pt}
    \centering
    \setlength{\abovecaptionskip}{0.1cm}
    \includegraphics[width=\linewidth]{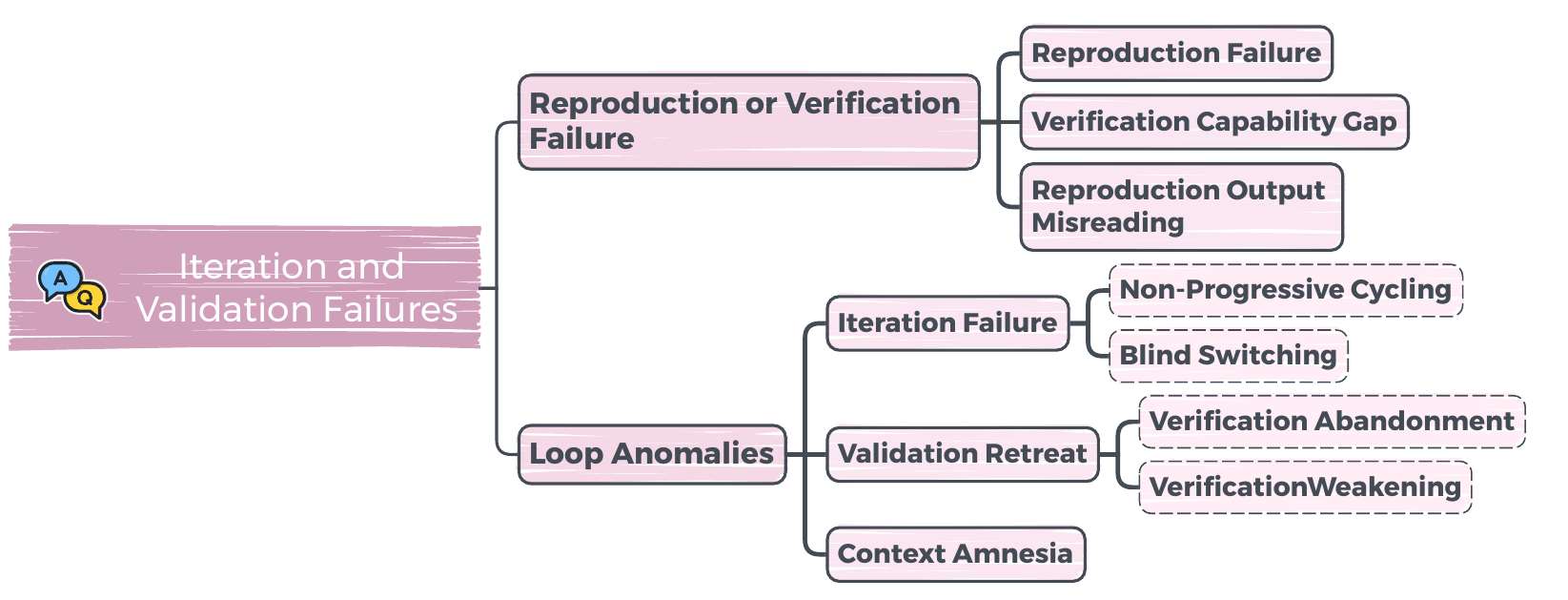}
    \caption{Failure modes in Iteration Verification stage.}
    \label{fig:mode3}
    \vspace{-10pt}
\end{wrapfigure}

Failure from this stage, particularly in agentic tools, involves the loop of testing a patch, interpreting the feedback an agent receives from both its tool interactions and test executions, and refining the solution. Figure~\ref{fig:mode3} illustrates a taxonomy of such failure modes.

\noindent\textbf{\textit{C1: Reproduction or Verification Failure.}}
This arises when the tool fails to set up a valid testing environment or correctly interpret its results.

\noindent\textit{C1.1: Reproduction/Validation Run Failure.} 
The tool often struggles to reproduce or validate issues because it mishandles project-specific environments and test setups. 
The tool's reproduction attempts may prune away essential components (\textit{e.g.}, database backends, configuration files), causing tests to fail for setup reasons.
For example, in frameworks like Django or Sphinx, failures commonly arise from missing or misconfigured dependencies, incorrect invocation of test runners, or misinterpretation of project-specific scripts and settings. 

\noindent\textit{C1.2: Insufficient Verification Capability.} This occurs when the tool's generated tests fail to fully validate a fix. For example, with visual outputs such as Matplotlib plots or web renderings, it may only check that a function runs or an image is produced, without verifying output correctness.
    
\noindent\textit{C1.3: Reproduction Output Misreading.} The tool sometimes misinterprets test feedback. For instance, even when failures are explicit or outputs unchanged, it may wrongly conclude success, leading to premature submission of broken or incomplete patches while the issue remains unresolved.

\noindent\textbf{\textit{C2: Iteration Anomalies.}}
This failure often arises when the tool's reasoning within the iterative loop becomes dysfunctional, hindering further progress toward a correct solution.

\noindent\textit{C2.1: Non-Progressive Iteration.} 
The tool may get trapped in repetitive modification cycles, repeatedly editing the same code fragment or configuration file across multiple interaction rounds without yielding meaningful progress. This often manifests as near-identical patches being regenerated with only trivial or cosmetic differences, such as variable renaming or shifting conditionals, while the underlying defect remains unaddressed. 

\noindent\textit{C2.2: Blind Strategy Switching.}
After a failed attempt, the tool may abruptly abandon its approach and switch to an unrelated strategy, yielding fragmented fixes across different code regions or incompatible patches. Lacking continuity, it fails to refine cumulatively and instead oscillates between disconnected changes.

\noindent\textit{C2.3: Validation Retreat.}
To make a test pass, the tool may modify the test case itself—for instance, by commenting out a failing assertion, loosening condition, or adjusting the test setup to accommodate its flawed patch—rather than correcting the underlying code.
For example, in \texttt{django\_\_django-16454}, after its initial patch failed the tests, the tool modified the test script instead of debugging, arguing it would ``better match real Django usage''. This merely tailored the test to its faulty code rather than meeting the original requirements.

\noindent\textbf{\textit{C3: Context Amnesia.}}
This failure occurs when the agent loses its original objective or its understanding of the code's current state during a long interaction \cite{liu2023lost}. In \texttt{django\_\_django-17084}, the agent initially identified a \texttt{GroupingError} but, after an unrelated setup error, lost its objective and wrongly shifted focus. More often, it fails to track its own edits, \textit{e.g.}, attempting changes with strings from the original file.

\section{Analysis of Failure Modes in LLM-based Issue Solving (RQ3)}\label{sec:rq3}
In this RQ, we analyze the distribution of failures across tools and task difficulty levels, uncover their root causes, and assess their downstream impact on the problem-solving process.

\begin{figure}[t] 
    \centering
    \setlength{\abovecaptionskip}{0.1cm}
    \includegraphics[width=0.9\linewidth]{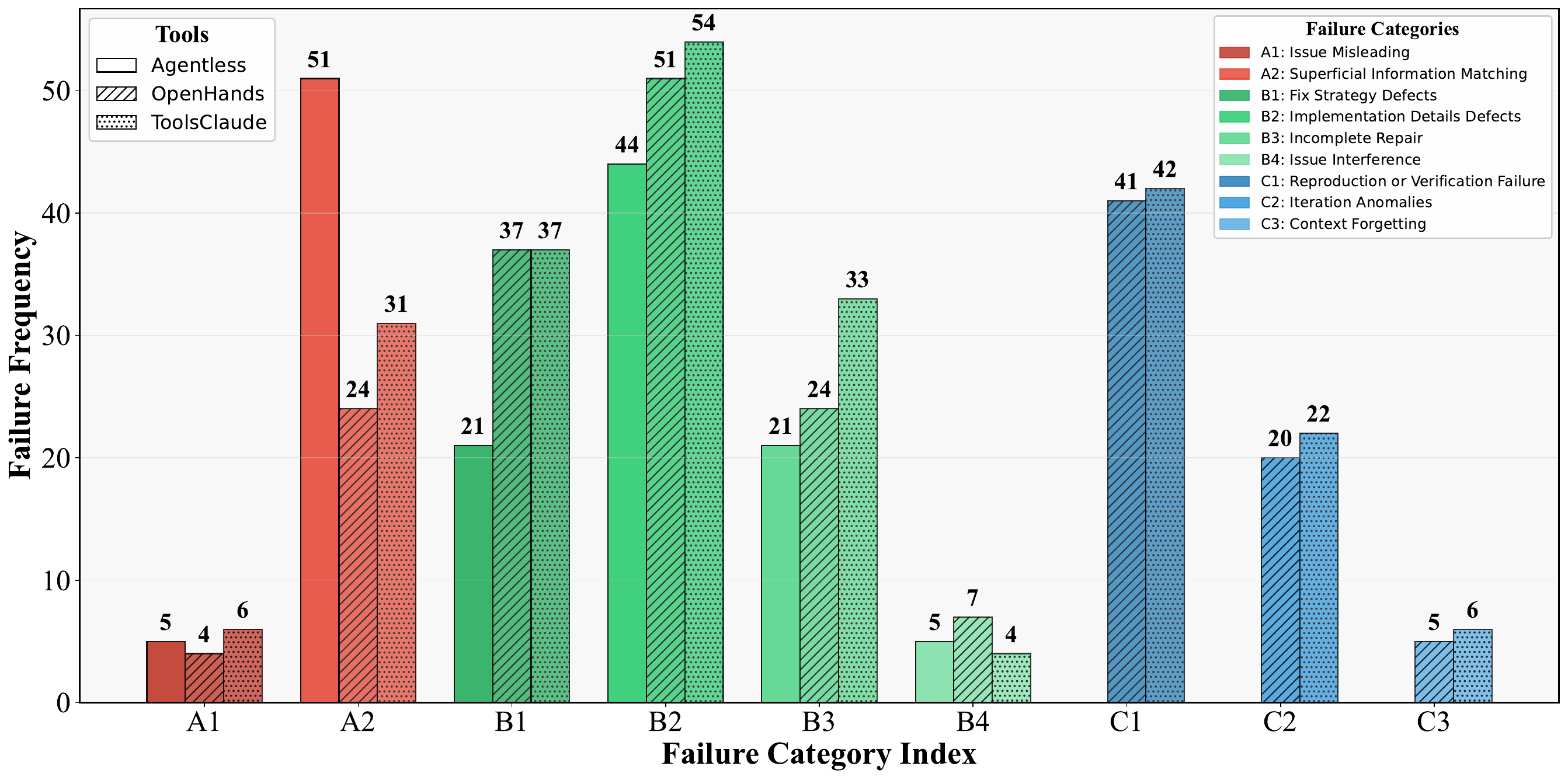}
    \caption{Distribution of failures across different tools.}
    \label{fig:frameworks_comparison}
    \vspace{-0.3cm}
\end{figure}

\subsection{Failure Distribution}

\begin{wrapfigure}{r}{0.55\textwidth}
\vspace{-0.3cm}
    \setlength{\abovecaptionskip}{0.1cm}
    \centering
    \includegraphics[width=\linewidth]{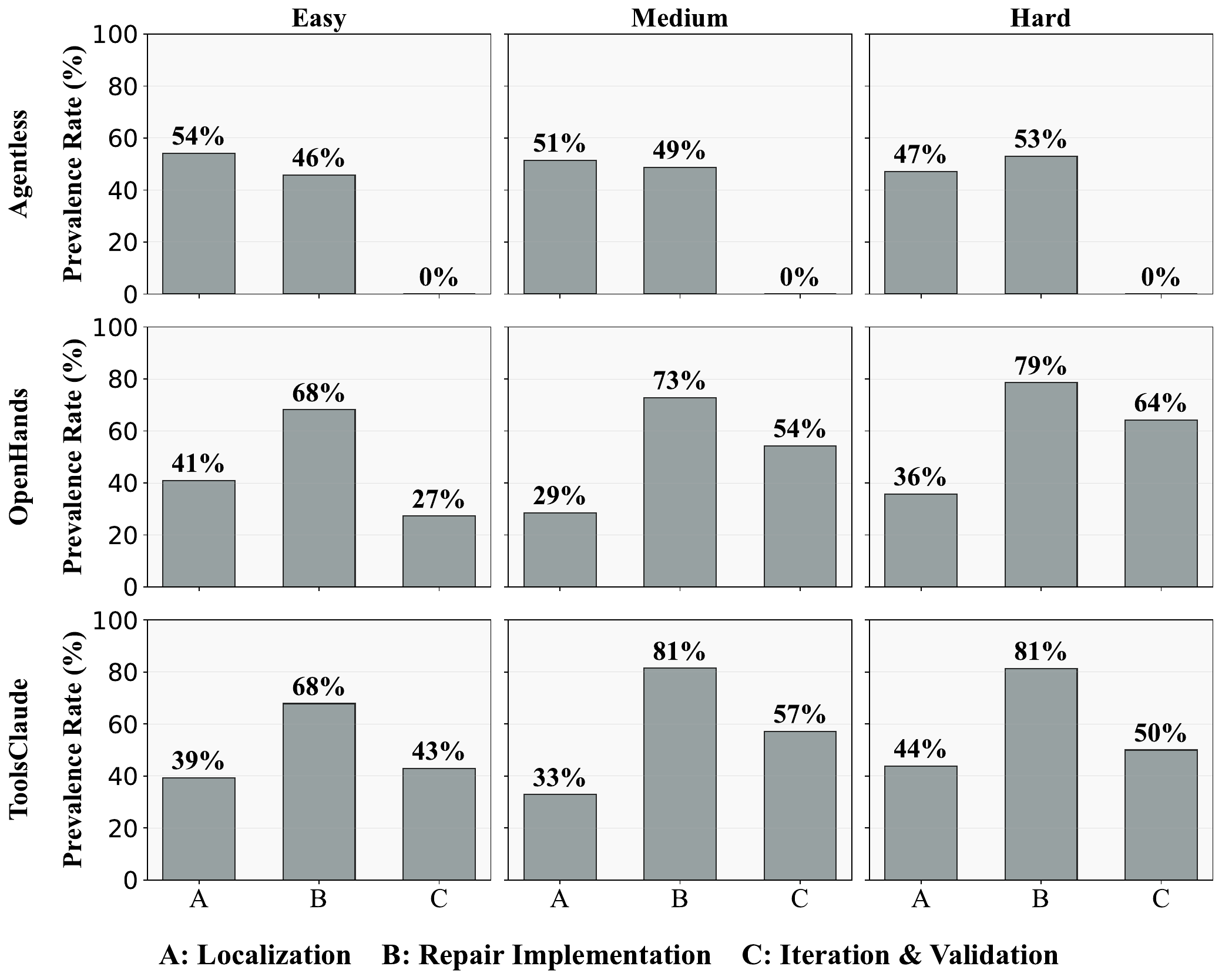}
    \caption{Failure stage distribution across task difficulty levels.}
    \label{fig:complexity_distribution}
    \vspace{-0.3cm}
\end{wrapfigure}

Figure~\ref{fig:frameworks_comparison} presents the distribution of primary failure modes across the three studied tools. Overall, the comparison highlights that the distribution of these failures is closely correlated with the tool's architecture. Specifically, each architecture exhibits a distinct ``failure fingerprint''. The pipeline-based Agentless is defined by its brittleness in the \textit{Localization} stage, with over half (\textbf{51.3\%}) of its failed issues containing a localization error. In contrast, agentic tools shift the failure burden to the \textit{Iteration \& Validation} stage, a phase entirely absent in Agentless, where roughly half of their failed issues contain errors (\textbf{49.1\%} for OpenHands and \textbf{51.7\%} for ToolsClaude).

We further analyze how failures at different stages (\textit{i.e.}, A: Localization, B: Repair, C. Iterative Verification) are distributed across task difficulty levels, with the results shown in Figure~\ref{fig:complexity_distribution}.
% We can observe a significant trend at the stage level.
On Easy tasks, a notable portion of failures still occur in the \textit{Localization} stage (41\% for OpenHands, 39\% for ToolsClaude), suggesting agents can be misled even on simple problems. However, as tasks progress to Medium and Hard, the prevalence of these localization failures generally drops. Concurrently, the proportion of failures in the \textit{Iteration \& Validation} stage rises dramatically, increasing from 27\% on Easy tasks to 64\% on Hard tasks for OpenHands. This pattern indicates that as tasks become more complex, the richer context provided may aid in localization, shifting the challenge from finding the bug to fixing it through a complex iterative process.

Besides, the rise in \textit{Iteration \& Validation} failures on more complex problems is driven by the increased prevalence of errors that test an agent's deeper cognitive abilities. For instance, failures in \textit{Algorithmic Implementation} (70.8\% in Medium tasks) and \textit{Logic Coordination} (72.3\% in Medium tasks) become dominant. Furthermore, on the most convoluted Hard tasks, we observe a notable presence of errors related to long-term strategic coherence, such as \textit{Context Forgetting} and \textit{Blind Strategy Switching}.

\vspace{1mm}
\begin{custommdframed}
\textbf{Finding 4:} 
Pipeline-based tools mainly fail in localization stage, whereas agentic tools become more prone to iteration anomalies failures as tasks grow harder. Yet across all settings, fix implementation remains the dominant bottleneck.
\end{custommdframed}
\vspace{1mm}

\subsection{Causes and Downstream Impacts of Failures}

As mentioned in \S\ref{sec:manual_analysis}), during the manual analysis, the annotators were instructed to document both the underlying causes and the observable effects of the identified failures. Building on the distribution patterns identified above, along with the labeled causes and effects, we now analyze why these failures emerge (\textbf{causes}) and examine their \textbf{downstream impacts}.

\subsubsection{Causes}
Our manual analysis results suggests that most failures are not simply the result of insufficient context or missing information. Rather, they reflect deeper flaws in the agent's reasoning process. To make sense of these flaws, we distill them into three overarching categories.
The most dominant cause (approx. 65\% of cases) is \textbf{flawed reasoning}, where the tool's internal reasoning logic leads it astray. This is the origin of the most stubborn failures observed in our study. For example, one common manifestation of flawed reasoning is an over-reliance on shallow heuristic—such as fixating on keywords from the issue description—which directly drives the catastrophic \textit{A2: Superficial Matching} failures. More importantly for agentic tools, \emph{getting stuck in a repetitive loop} is the direct cause of \textit{C2.1: Non-progressive Iteration}. This pattern, where an agent becomes fixated on a failing approach, highlights a core vulnerability: the lack of an effective mechanism for self-correction. This inability to pivot from a failing strategy leads to cognitive deadlocks. The second cause is \textbf{knowledge deficiency} (approx. 25\%),
which arises either from missing codebase-specific context (resulting in \textit{B3: Incomplete Repair}) or from outdated domain expertise (\textit{B2.3: Insufficient Domain Knowledge}).
Finally, \textbf{environmental friction} (approx. 10\%) describes the agent's failure to properly interact with its environment, which is the cause of verification issues or tool usage errors.

\subsubsection{Downstream Impacts}
The downstream impacts of the failures are significant resource consumption and fruitless explorations.
In particular, failures caused by \textit{flawed reasoning} typically
drive agents into persistent loop anomalies, where failed issues involve on average \textit{3.5 times} more interaction steps than successful ones. 
Similarly, failures derived from \textit{knowledge deficiency} leads to prolonged but misguided searches through the codebase. Failures caused by \textit{environmental friction} also proves costly, especially in complex repositories like \texttt{django} or \texttt{sphinx}, where agents expend numerous interaction rounds merely attempting to reproduce the issue, often before the core repair task can even begin.

\vspace{1mm}
\begin{custommdframed}
\textbf{Finding 5:} 
Most failures in agentic tools arise from flawed reasoning. Single agent often gets stuck in cognitive deadlocks, such as persisting with a failed strategy, and lack mechanisms for strategic self-correction. This makes the single-agent paradigm inherently vulnerable, underscoring the need for external oversight.
\end{custommdframed}
\vspace{1mm}

\section{Mitigation Exploration with a Collaborative Architecture (RQ4)}\label{sec:rq4}

Our findings in RQ3 (\S\ref{sec:rq3}) reveal that most agent failures arise from \textbf{flawed reasoning}, which often leads to cognitive deadlocks. Furthermore, single agent lacks effective self-correction mechanisms to escape these unproductive loops \cite{liu2025breaking}. Building on these insights, we introduce a collaborative framework, the \textbf{Expert-Executor Model}, designed to emulate human peer review to mitigate such failures. We implemented this architecture based on the open-source OpenHands tool.

\subsection{The Expert-Executor Model}
Our framework comprises two distinct but deeply integrated agents:
\begin{itemize}[leftmargin=*]
    \item \textbf{Execution Agent:} An agent responsible for solving issue tasks through code analysis, patch generation, and test execution. Its prompt explicitly requires it to consult the Expert Agent at key decision points—for example, after exploring the repository to validate its understanding of the issue, or before implementing a fix to have its proposed solution reviewed.
    \item \textbf{Expert Agent:} A specialized LLM agent acting as a senior technical lead. Its primary function is to monitor the Execution Agent's workflow, identify potential strategic flaws, and provide corrective guidance. Based on our analysis of failure modes in RQ3 (Section \S\ref{sec:rq3}), we structure its prompt around our failure taxonomy (developed in Section \S\ref{sec:taxonomy}). The prompt provides the name and a detailed explanation for each failure mode. The detailed prompt template can be found in our replication package.\footnote{\url{https://anonymous.4open.science/r/IssueSolvingEmpirical/Expert-Executor/evaluation/benchmarks/swe_bench/EXPERT_HYBRID_PROMPT.md}}
\end{itemize}

To solve an issue, the process begins with the \textbf{Execution Agent}, which first reads the issue description, analyzes the relevant code, proposes candidate patches, and executes tests to validate them. Unlike single agent tools (such as the two tools analyzed in our study), the \textbf{Execution Agent} does not work in isolation. Instead, it collaborates closely with the \textbf{Expert Agent}, which serves as a strategic reviewer to prevent reasoning deadlocks and guide the solution process toward correctness and efficiency. This collaboration is achieved through two primary forms of interaction:
\begin{itemize}[leftmargin=*]
    \item \textbf{Active Consultation:} The Execution Agent is prompted to actively seek guidance at specific times. For instance, before exploring the repository, it shares its understanding of the problem with the Expert Agent. Similarly, before writing the code for a fix, it presents its proposed solution for validation. It is also required to ask for help if it gets stuck. 
    
    \item \textbf{Passive Review:} The Expert Agent periodically reviews the session history to detect flawed patterns like stagnation or strategic errors. If such issues are found, it autonomously intervenes with course-correcting advice to counteract silent failures.
\end{itemize}

\subsection{Experimental Setup and Result Analysis}

We implement our model based on \textbf{OpenHands}, as it represents a state-of-the-art, open-source agentic tool.
% whose codebase could be directly adapted to realize our collaborative design. 
To evaluate the effectiveness of our tool, we conduct experiments on the 108 issues from our annotated dataset on which the baseline OpenHands agent had failed. Our goal was to determine if collaborative oversight could resolve these previously intractable cases. The model successfully resolved \textbf{24 of these 108} failed issues, achieving a \textbf{22.2\%} success rate on a challenging set of tasks where a state-of-the-art single agent had already failed. In our Expert–Executor architecture, both the Execution Agent and the Expert Agent leverage \textbf{Claude 3.5 Sonnet}. Additionally, the interval count for \textbf{Passive Review} is set to 25, with the total number of rounds configured to 40. For comparison, a single-agent baseline (\textbf{OpenHands + Claude 4 Sonnet}) resolved only 7 of these cases, underscoring the effectiveness of collaborative oversight.

\begin{wrapfigure}{r}{0.6\textwidth}
    \vspace{-10pt}
    \setlength{\abovecaptionskip}{0.1cm}
    \centering
    \includegraphics[width=\linewidth]{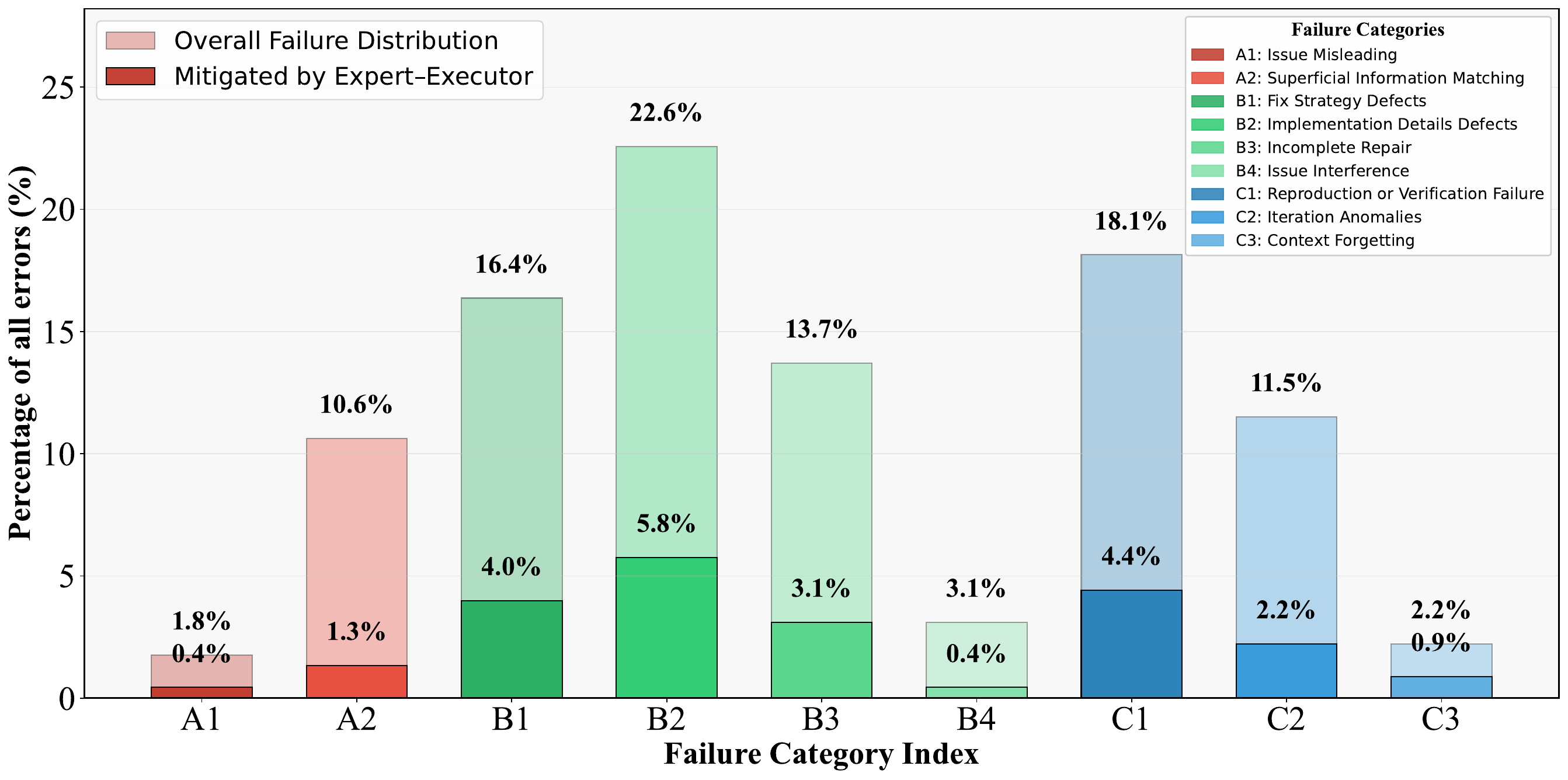}
    \caption{Failure distribution comparison between OpenHands and our Expert-Executor model.
    }
    \label{fig:error_distribution}
    \vspace{-10pt}
\end{wrapfigure}

Figure~\ref{fig:error_distribution} compares the baseline failure distribution of the OpenHands agent with the subset of failures successfully resolved by our Expert–Executor model, highlighting both the dominant error categories and the targeted mitigation achieved through collaborative oversight. The figure shows our Expert-Executor model demonstrates substantial mitigation in dominant failure categories, particularly in addressing \textit{Fix Strategy Defects} (\textit{B1}), producing correct code for \textit{Implementation Details Defects} (\textit{B2}), and accurately interpreting test feedback for \textit{Reproduction or Verification Failures} (\textit{C1}).

\begin{wrapfigure}{r}{0.6\textwidth}
    \setlength{\abovecaptionskip}{0.1cm}
    \centering
    \includegraphics[width=\linewidth]{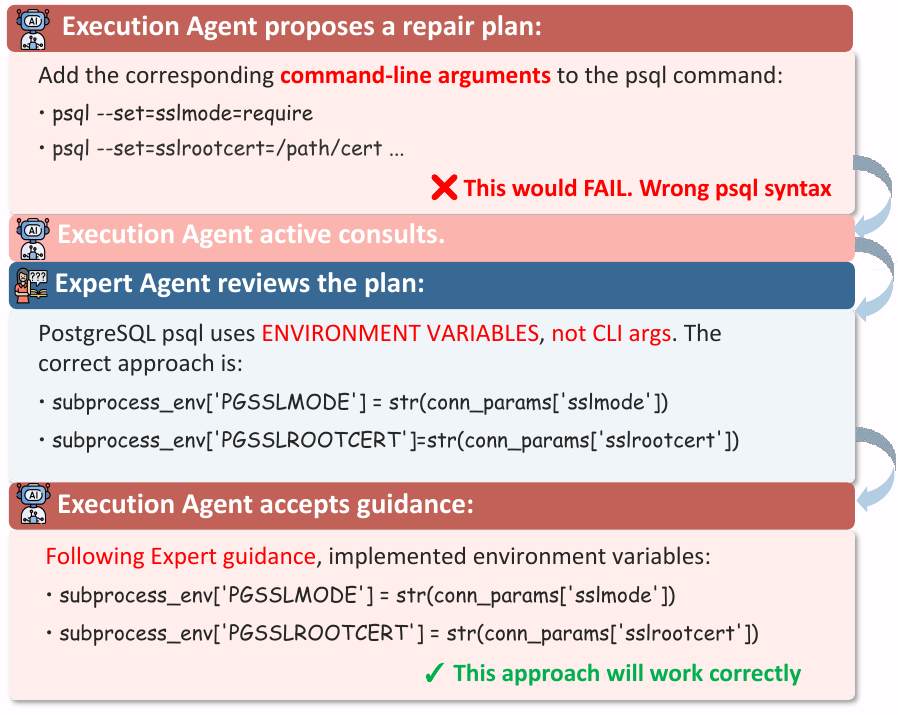}
    \caption{Example of issue solving with Expert-Executor Model.
    }
    \label{fig:expert-example}
    \vspace{-10pt}
\end{wrapfigure}

Our analysis of the 24 newly solved instances further provides direct evidence of our model's targeted mitigation against these dominant failure types. Specifically, the Expert's interventions proved most effective against the very patterns that plagued the baseline, successfully correcting the most common iterative error, \textit{Reproduction Output Misreading} (in 6 solved cases), and the most frequent strategic and implementation flaws, including \textit{Evasive Repair} (5 cases) and \textit{Control Flow Defects} (5 cases). By succeeding on previously failed tasks, including 16 Medium and 1 Hard-difficulty issues, our Expert-Executor model validates its potential to resolve not just simple bugs, but complex reasoning failures.

For example, as shown in Figure~\ref{fig:expert-example}, in \texttt{django\_\_django-11239}, when tasked with adding SSL support to Django's \texttt{dbshell}, the Executor agent correctly identifies the goal but proposed a failed implementation using command-line arguments (\textit{e.g.}, \texttt{--set=sslmode=...}) for the PostgreSQL client. As required by the prompt, the Executor engages in \textbf{Active Consultation}, presenting its plan to the Expert Agent for validation \textit{before} implementing the fix. The Expert then provides the critical correction, explaining that \texttt{psql} requires environment variables (\textit{e.g.}, \texttt{PGSSLMODE}) for SSL configuration. This pre-emptive review, initiated by the Executor itself, prevents a predictable failure and guided it directly to the correct.

\vspace{1mm}
\begin{custommdframed}
\textbf{Finding 6:} 
Our proposed Expert-Executor model mitigates core reasoning failures of single agents through strategic oversight and peer-review-like collaboration. By breaking cognitive deadlocks and correcting flawed strategies, it turned 22.2\% of previously unsolvable problems into solvable ones, showing the effectiveness of an external verification layer in building more robust autonomous agents.
\end{custommdframed}

\section{Discussion}

\subsection{Implications}

\noindent\textbf{Implications for Researchers.} Current benchmarks \cite{jimenez2024swe-bench} for evaluating issue solving capabilities, dominated by issue-solving rates, obscure critical failure patterns (Finding 4). To foster genuine progress, evaluations should adopt diagnostic benchmarking with failure taxonomies, such as the one we propose, to reveal evaluated tools' specific weaknesses and guide targeted improvements beyond a single success score. Our analysis further highlights cognitive deadlocks as a central failure mode, which scaling monolithic models alone cannot resolve (Finding 5). Addressing this requires a shift toward architectures that explicitly mitigate reasoning flaws—most notably through \textbf{collaborative multi-agent tools} that introduce strategic oversight (Finding 6) and \textbf{self-correcting agents} \cite{madaan2023self, shinn2023reflexionlanguageagentsverbal,chen2023teaching} capable of recognizing and escaping unproductive loops (Finding 3). Together, these directions point to a research agenda that moves beyond raw scale toward systems designed for diagnosis, collaboration, and adaptive reasoning.

\noindent\textbf{Implications for Practitioners.} No single architecture excels at all tasks (Finding 1). Practitioners should adopt a portfolio approach: pipeline-based tools are well-suited for simple, localized bugs, while more resilient agentic tools are preferable for complex, multi-file problems (Finding 2). Recognizing each tool's \textit{failure fingerprint} enables more informed choices and better anticipation of likely failure points. At the same time, agents often become trapped in repetitive or unproductive loops (Finding 3). Rather than letting them run indefinitely, practitioners should monitor their progress and \textbf{intervene when they are clearly off-track}. Indeed, the principle of leveraging human insight to overcome automated limitations is well-established in program repair \cite{geethal2023human}. Timely, high-level guidance \cite{böhme2019humanintheloopautomaticprogramrepair} to a senior developer can break deadlocks and save significant time and resources.

\subsection{Threats to Validity}
\noindent\textbf{Threats to external validity} relate to the generalizability of our constructed failure mode taxonomy and findings. Our study is grounded in SWE-Bench-Verified, spanning tasks from 12 diverse Python repositories, and examines three tools (OpenHands-CodeAct-2.1, Tools + Claude 3.5 Sonnet, Agentless-1.5) that represent distinct architectural paradigms. This paradigm-level focus suggests that the identified ``failure fingerprints'' extend beyond specific implementations. Our taxonomy reached \textit{theoretical saturation} during annotation, indicating coverage of the relevant failure space. Still, the analysis is limited to Python and a single backbone LLM (Claude-3.5 Sonnet). Future work should test its applicability to other languages and LLMs \cite{zan2025multiswebenchmultilingualbenchmarkissue}.

\noindent\textbf{Threats to internal validity} relate to the subjectivity of our manual analysis, as different annotators may interpret the same execution trace differently. To mitigate this, we designed a multi-stage protocol: \ding{172} a detailed codebook based on a 50-case pilot; \ding{173} a custom web platform for standardized annotation; \ding{174} double-coding of each instance by at least two of four annotators; and \ding{175} structured disagreement resolution with senior arbitration. Inter-annotator agreement, measured by Cohen's Kappa (0.72–0.77) \cite{P_rez_2020}, indicates substantial reliability, underscoring the robustness of our taxonomy.

\section{Conclusion}
In this paper, we present the first in-depth empirical study on why LLM agents fail at automated issue solving tasks. We introduce a comprehensive taxonomy of failure modes comprising
3 primary phases, 9 main categories, and 25 fine-grained subcategories. Our further analysis of these failure modes reveals that different agentic architectures exhibit distinct failure fingerprints: pipeline-based tools are brittle and fail early in Localization stage, while agentic tools succumb to late-stage Iteration loops caused by flawed reasoning and cognitive deadlocks. To mitigate this, we propose a collaborative Expert-Executor framework that mimics human peer review. This model successfully resolved \textit{22.2\%} of previously intractable issues by breaking these reasoning deadlocks, demonstrating a promising path toward building more robust agents for automated issue solving.

\section*{Data Availability}
To facilitate further research, we provide the replication package at \url{https://anonymous.4open.science/r/IssueSolvingEmpirical}.

\bibliographystyle{ACM-Reference-Format}
\bibliography{ref}

\end{document}